\newif\ifsingle

\ifsingle
\documentclass[11pt,draftclsnofoot, onecolumn]{IEEEtran}		
\else		
\documentclass[10pt,final, twocolumn]{IEEEtran}
\fi

\usepackage{times}
\usepackage{amsmath,dsfont}
\usepackage{amssymb,amsthm}
\usepackage{epsfig,verbatim}
\usepackage{subfigure}
\usepackage{setspace}
\usepackage{color}
\usepackage{cite}
\usepackage{epstopdf}
\usepackage{graphics}
\usepackage{accents}
\usepackage{acronym}
\usepackage[bookmarks,colorlinks]{hyperref}
\usepackage{booktabs}
\usepackage{mathtools}
\usepackage{enumitem}
\usepackage{Quant}
\usepackage{bm}
\usepackage{multirow}

\usepackage[ruled,linesnumbered,vlined]{algorithm2e}
\SetKwInput{KwData}{\textbf{Init}} 
\let\oldnl\nl
\newcommand{\nonl}{\renewcommand{\nl}{\let\nl\oldnl}}

\newtheorem{definition}{Definition}

\newcommand{\myVec}[1]{{\boldsymbol{#1}}}

\newcommand{\mySet}[1]{\mathcal{#1}}
\newcommand{\topen}{t_i^{\rm o}}
\newcommand{\tclose}{t_i^{\rm c}}

\acrodef{bb}[BB]{Bollinger Bands}
\acrodef{ma}[MA]{Moving Average}
\acrodef{std}[STD]{Standard Deviation}
\acrodef{op}[OP]{Open Position}
\acrodef{cp}[CP]{Closed Position}
\acrodef{hp}[HP]{Hold Position}
\acrodef{pnl}[PNL]{Profit and Loss}
\acrodef{rl}[RL]{Reinforcement Learning}
\acrodef{kf}[KF]{Kalman Filter}
\acrodef{ss}[SS]{Space State}
\acrodef{ci}[CI]{Co-Integration}
\acrodef{pci}[PCI]{Partially Co-Integration}
\acrodef{knet}[KNET]{KalmanNet}
\acrodef{ddqn}[DDQN]{Double Deep Q Network}
\acrodef{mb}[MB]{model-based}
\acrodef{dd}[DD]{data-driven} 
\acrodef{kbpt}[KBPT]{KalmenNet-aided Bollinger bands Pairs Trading}

\acrodef{mse}[MSE]{mean-squared error}
\acrodef{iot}[IOT]{Internet of Things}
\acrodef{rmse}[RMSE]{root mean squared error}
\acrodef{rmspe}[RMSPE]{root mean squared periodic error}
\acrodef{mmse}[MMSE]{{minimum mean-squared error}}
\acrodef{lmmse}[LMMSE]{{linear} MMSE}
\acrodef{mle}[MLE]{maximum likelihood estimation}
\acrodef{snr}[SNR]{signal-to-noise ratio}
\acrodef{dnn}[DNN]{deep neural network}
\acrodef{admm}[ADMM]{alternating direction method of multipliers}
\acrodef{dadmm}[D-ADMM]{distributed alternating direction method of multipliers}

\IEEEoverridecommandlockouts
\title{
Neural Augmented Kalman Filtering with Bollinger Bands for Pairs Trading}

\author{
\IEEEauthorblockN{Amit Milstein, Haoran Deng,  Guy Revach, Hai Morgenstern, and Nir Shlezinger
\thanks{
Parts of this work were accepted for presentation in the 2023 IEEE International Conference on Acoustics Speech, and Signal Processing (ICASSP) as the paper \cite{deng2023kalmanbot}. 
A. Milstein and N. Shlezinger are with the School of ECE, Ben-Gurion University of the Negev, Israel (e-mail: amitmils@post.bgu.ac.il; nirshl@bgu.ac.il).  		H. Deng and G. Revach are with the Institute for Signal and Information Processing, D-ITET, ETH Zürich, 
		Switzerland (e-mail: haodeng@student.ethz.ch; grevach@ethz.ch). 
		H. Morgenstern is unaffiliated (e-mail: hai.morgenstern@gmail.com). 
}}}

\begin{document}

\maketitle
\pagestyle{plain}
\thispagestyle{plain}

%
\begin{abstract}
Pairs trading is a family of trading techniques that determine their policies based on monitoring the relationships between pairs of assets. A common pairs trading approach relies on describing the pair-wise relationship as a linear \ac{ss} model with Gaussian noise. This representation  facilitates extracting financial indicators  with low complexity and latency using a \ac{kf}, that are then processed using classic policies such as \ac{bb}. However, such \ac{ss} models are inherently approximated and mismatched, often degrading the revenue. In this work, we propose \ac{kbpt}, a deep learning aided policy that augments the operation of \ac{kf}-aided \ac{bb} trading. \ac{kbpt} is designed by formulating an extended \ac{ss} model for pairs trading that approximates their relationship as holding partial co-integration. This \ac{ss} model is utilized by a trading policy that augments \ac{kf}-\ac{bb} trading with a dedicated neural network based on the \acl{kn} architecture. The resulting \ac{kbpt} is trained in a two-stage manner which first tunes the tracking algorithm in an unsupervised manner independently of the trading task, followed by its adaptation to track the financial indicators to maximize revenue while approximating  \ac{bb}  with a differentiable mapping. \ac{kbpt} thus leverages data to overcome the approximated nature of the \ac{ss} model, converting the \ac{kf}-\ac{bb} policy into a trainable model. We empirically demonstrate that our proposed \ac{kbpt} systematically yields improved revenue compared with model-based and data-driven benchmarks over various different assets.  
\end{abstract}
\acresetall

\section{Introduction}\label{sec:intro}
Quantitative methods constitute the fundamental mathematical framework for  analysis and prediction in financial markets~\cite{markowitz1991foundations, fama1992cross}. A common type of quantitative methods is algorithmic trading~\cite{chan2013algorithmic}, which deals with decision-making carried out by an agent (i.e., a trader) for the purpose of maximizing a cumulative reward,  most commonly achieving a  high \ac{pnl} balance in the market. Quantitative trading schemes are typically comprised of two main stages: the agent first tracks a stochastic process that describes the prices of the assets of interest in order to extract useful trading indicators. Then, these financial indicators are used as a basis for decision making by setting a trading policy~\cite{akansu2016financial,de2018advances,Algo_trading}. 
%

Quantitative trading requires a decision making mechanism given application time constraints, i.e., a trading policy that outputs a position based on the trading indicators.
Such policies are typically based on indicators obtained as statistical predictions of an asset price~\cite{jegadeesh1993returns}. A popular classical policy is the \emph{\ac{bb}} \cite{bollinger1992using}, which is based on the intuition that if the price is much less than its mean, it will rise back to normal level and thus one should long this asset. Due to the fact that this method is not linear, it hedges the risk by constraining the investment.

Classical trading schemes such as \ac{bb} work well for single \emph{stationary} (and specifically, \emph{mean-reverting}) processes~\cite{poterba1988mean}. It is therefore sought-after to look for stationary assets, though some schemes only look for the weaker condition of mean reverting, e.g., using the Ornstein–Uhlenbeck formula~\cite{maller2009ornstein}.
Accordingly, algorithmic tracking of financial  processes is typically based on imposing a model on their temporal evolution~\cite{benidis2018optimization}. A common approach imposes simple linear stochastic stationary model~\cite{tsay2005analysis}, often based on autoregressive and moving average models~\cite{feng2016signal}. While assets are rarely stationary in real markets, their differences and {\em spread} (i.e., linear combination) are in some cases faithfully captured as being stationary, and thus such techniques are commonly adopted in {\em pairs trading}~\cite{elliott2005pairs, chan2013algorithmic}. The spread evolution  and its relationship with the assets pair is often described using a \ac{ss} model~\cite{puspaningrum2012pairs,clegg2018pairs,krauss2017statistical}, enabling tracking with a \ac{kf}~\cite[Ch. 10]{feng2016signal}.
 A core challenge with combining financial policies with algorithmic tracking based on such statistical models it that they typically require strong assumptions and prior financial knowledge. For instance, to utilize the \ac{kf} for spread tracking, one has to faithfully capture the  pairs trading as a linear Gaussian \ac{ss} model. Such models often fail to capture complicated patterns of real world financial assets, which in turn leads to poor trading  policies.

To overcome the drawbacks of classic model-based methods, recent years have witnessed a growing interest in the use of model-agnostic deep learning.
 Deep learning systems are used to capture the time evolution of financial assets \cite{lim2021time}, extract features  for trading \cite{zhang2018improving}, and determine trading policies~\cite{spooner2018market}, see survey in \cite{benidis2022deep}.
Common deep learning architectures for financial modelling and prediction include \acp{rnn}~\cite{xu2018stock}, auto-encoders~\cite{bao2017deep}, anomaly detection~\cite{ahmed2016survey} and attention models~\cite{DBLP:conf/ijcai/QinSCCJC17,zhu2021attention}.  \ac{rl} is considered for training deep trading policies \cite{DBLP:conf/nips/MoodyS98, sun2023reinforcement, borodin2003can, brim2020deep, spooner2018market,han2023mastering} to maximize the reward in an \acl{e2e} fashion.  In order to generate various inputs,  it was proposed to use deep learning based natural language processing to analyze social media and news for trading \cite{xu2018stock,zhang2018improving}.
Despite their growing popularity, deep learning based quantitative methods are subject to several  drawbacks. They are based on highly parameterized black boxes, giving rise to latency considerations. Moreover, deep learning based policies lack the interpretability and reliability of model-based methods, and do not incorporate established  models which is core in pairs trading. In addition, these methods tend to have a long training time and require large volumes of data for training, which can constitute a limiting factor in high-frequency trading. This motivates designing trading techniques that simultaneously benefit from the approximated modelling adopted by classical trading schemes alongside the abstractness and capabilities of data-driven deep learning methods. 

In this work, we propose \ac{kbpt}, a pairs trading algorithm that combines \ac{ss} model-based trading policies with deep learning tools, based on {\em model-based deep learning} methodology~\cite{shlezinger2020model, shlezinger2022model,shlezinger2023model}. \ac{kbpt} is derived by proposing a novel \ac{ss} model representation for pairs trading obtained from assuming partial co-integration~\cite{clegg2018pairs} combined with an autoregressive prior imposed on the spread.
%
As opposed to previous \ac{ss} model-based trading policies that utilize, e.g., \ac{kf} with \ac{bb} for setting the position, thus implicitly assuming that the \ac{ss} model is Gaussian and accurate, we design our policy to particularly cope with the approximated nature of the \ac{ss} model and its expected non-Gaussianity. This is achieved by having \ac{kbpt}  preserve the flow of \ac{kf}-\ac{bb} trading, retaining its structured modeling and interpretability, while augmenting the \ac{kf} with a trainable \ac{rnn} following the recently proposed \acl{kn}~\cite{revach2022kalmannet}. The resulting neural augmentation, in which the specific computation of the \ac{kf} that depends on the underlying stochasticity is learned, leverages data to track the spread in \emph{partially} known and {\em non-Gaussian} \ac{ss} models.

We propose a dedicated training scheme for \ac{kbpt} that learns the pairs trading policy from sequences of past assets pairs. The learning method is based on a two-stage procedure, where we first train \acl{kn} separately from the trading task as a form of pretraining. There, we overcome the fact that there is no ground-truth spread value by leveraging the interpretable architecture of \acl{kn}, and particularly its internal prediction of the next observation which follows from the \ac{kf} flow, for unsupervised learning~\cite{revach2021unsupervised}. Then, we train the overall trading policy, combining the neural augmented \acl{kn} with a customized \ac{bb} mapping that is differentiable, such that the tracking algorithm learns to produce features that are most useful in the sense of maximizing the \ac{pnl} rather than accurately tracking the prices.
By that, we gain the ability to cope with modeling mismatch, as the resulting architecture converts the model-based trading algorithm into a trainable discriminative model~\cite{shlezinger2022discriminative} that is trained end-to-end to maximize the \ac{pnl} as a cumulative reward. 

Our empirical study compares \ac{kbpt} with both model-based trading and with deep \ac{rl}-based policies for various assets pairs. There, we demonstrate the individual gains of each of the ingredients of \ac{kbpt}, including the usefulness of the extended \ac{ss} model underlying \ac{kbpt}, as well as  the superiority of the proposed hybrid algorithm in systematically achieving higher \ac{pnl} compared with all considered benchmarks. 
Our work extends upon its preliminary findings reported in~\cite{deng2023kalmanbot} in the proposal of the new partially co-integrated \ac{ss} model, the incorporation of a dedicated accumulated reward loss and the two-stage training methods, as well as in the extensive discussion, derivation, and experimental evaluations.

The rest of this paper is organized as follows: Section~\ref{sec:Preliminaries} covers  preliminaries in model-based trading and formulates the problem; Section~\ref{sec:State} describes the different \ac{ss} models in pairs trading and presents our proposed model; Section~\ref{sec:Nerual_Aug_TP} details our proposed hybrid \ac{kbpt} policy along with its learning procedure; Section~\ref{sec:emp_eval} presents the empirical study of \ac{kbpt}, contrasting it with both \acl{mb} and \acl{dd} policies; while Section~\ref{ssec:conclusions} provides concluding remarks.

Throughout this paper we use boldface lower-case letters for vectors; e.g., $\gvec{x}$, and boldface uppercase letters for matrices, e.g., for $\gvec{X}$. We denote the step function as $\unitstep{\cdot}$, with $\unitstep{t} =1$ for $t> 0$ and $\unitstep{t} =0$ for $t\leq0$, while $\mathbb{E}\{\cdot\}$ is the notation for stochastic expectation. 
We use the term {\em stationary process} to refer to a stochastic process that is stationary in the wide sense. For consistency, the prices of all assets is given in USD.

%
\section{Preliminaries and Problem Formulation}\label{sec:Preliminaries}
In this section we formulate the considered model for pairs trading. To that aim, we first review necessary preliminaries in quantitative trading  in Subsection~\ref{ssec:trading}, and recall the \ac{bb} policy in Subsection~\ref{ssec:Bollinger}. These preliminaries are then used to formulate the problem in Subsection~\ref{ssec:problem}.

\subsection{Trading Formulation}\label{ssec:trading} 
Trading strategies  refer to the determining of investment policies based on the monitoring of financial assets. Accordingly, trading  strategies can be generally divided into two stages: $(i)$ tracking of the assets into financial indicators; and $(ii)$ the trading policy that is based on these indicators~\cite{Algo_trading}.

\subsubsection{Tracking}
A crucial part of any trading scheme is constantly evaluating and analyzing the financial markets, individual securities, or sectors. Information such as  price movements, volatility, liquidity, volume, momentum, and market breadth is valuable for making informed decisions in the trading market. Using this financial data, one can derive financial indicators which enable the trader to get insight on potential entry and exit points, assess risks, and ultimately optimize the investment strategy. Quantitative financial indicators can include technical indicators (e.g., moving averages, relative strength index) \cite{morrison2012microeconomic}, fundamental indicators (e.g., earnings per share, price-to-earnings ratio), or macroeconomic indicators (e.g., GDP growth rate, inflation rate)~\cite{pilinkus2010macroeconomic}.

To formulate this mathematically, we use $\gscal{d}_t$ to denote the financial information (e.g., assets price) at time $t>0$. A \textit{financial tracker}, denoted $\varphi$, is a mapping of all the financial data accumulated until time $t$ into financial indicators $\gscal{z}_t$, i.e.,
\begin{equation}
\label{eq:trackingmap}
  \varphi:\{\gscal{d}_\tau\}_{\tau\leq t}\mapsto\gscal{z}_t.  
\end{equation}
The financial indicator should provide sufficient information for the  policy to dictate the current decision, as detailed next. 

\subsubsection{Policy}
The policy component of a trading scheme, denoted by $\pi$, refers to the rules, guidelines, and principles that govern the decision-making process and the execution of trades. The policy component in general may encompass both quantitative and qualitative aspects:
Quantitative aspects can involve specific parameters, thresholds, or algorithms based on financial indicators or other mathematical models. Qualitative aspects consider factors such as market conditions, investor sentiment, news events, or expert judgment. 

 The policy is the last step of the trading scheme and it outputs the recommended actions for the trader to take in order to optimize profits. We refer to the return of each trade transaction the {\em reward}. In quantitative trading, the action at time $t$, denoted $\myVec{p}_t$, is determined using a trading policy $\pi$ based on the current indicator $\gscal{z}_t$ as well as past actions and indicators, namely,
  \begin{equation}
\label{eq:policymap}
     \pi:\{\gscal{z}_{\tau}\}_{\tau\leq t},\{\myVec{p}_{\tau}\}_{\tau<t}\mapsto\myVec{p}_t.
 \end{equation}

 We henceforth focus on settings where 
 \begin{enumerate}[label={A\arabic*}]
     \item \label{itm:Asset} The  information $\gscal{d}_t$ represents the price of an asset.
     \item  \label{itm:Position} The actions correspond to long/short decisions on $\gscal{d}_t$, i.e., holding positive or negative quantities, respectively.
 \end{enumerate}
 The action space in \ref{itm:Position} indicates that $\myVec{p}_t$ encapsulates open and close decisions. We formulate this by writing  $\myVec{p}_t=[\gscal{op}_t,\gscal{cp}_t]$, where $\gscal{op}_t\in\set{-1,0,1}$ is the {\em open position policy} that signals if to short, hold or long the asset, respectively; and $\gscal{cp}_t\in\set{0,1}$ is the {\em close position policy}, which gets the value $1$ when an existing open position (e.g., from time $t-1$) needs to be closed. Otherwise, if a position needs to remain open or there is no open position, it gets the value of $0$. The order in which  positions are taken involves  first checking if the closing criteria is met,  and then checking whether to open one. We say that $\myVec{p}_t$ is an {\em active position} if $\gscal{op}_t = \pm1$.

 \subsubsection{Reward}
Under \ref{itm:Asset}-\ref{itm:Position}, one can mathematically formulate the reward accumulated for an active position. To that aim,  let $\topen$ be the time the $i$th active position is taken and $\tclose$  the time it is closed. Accordingly, the reward obtained for the $i$th activity of   of policy $\pi$ with financial tracker $\varphi$, denoted by $r^{\varphi,\pi}_i$, is computed based the difference in the asset price over the activity period and whether it was  long or short  via
 \begin{equation}
 \label{eq:reward}
   r^{\varphi,\pi}_i = \gscal{op}_{\topen} \cdot (\gscal{d}_{\tclose} - \gscal{d}_{\topen}).
 \end{equation} 
The reward in \eqref{eq:reward} can be positive or negative, i.e., profit or loss, respectively.

%
\subsection{Bollinger Bands Trading Policy}\label{ssec:Bollinger}
\begin{figure}
  \centering
  \includegraphics[width=0.5\textwidth]{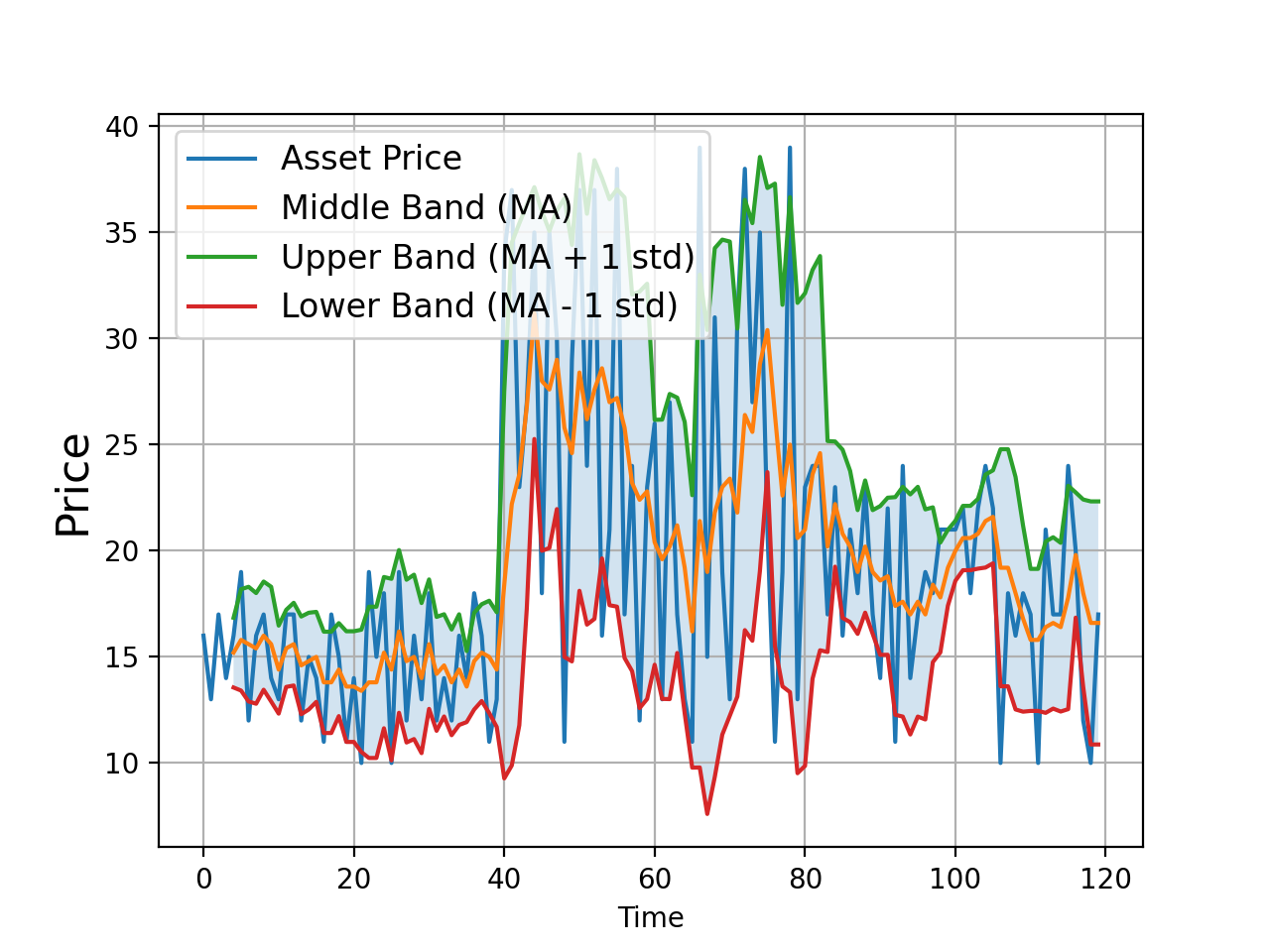}
  \caption{Asset price with Bollinger Bands illustration}
  \label{fig:bb_figure}
\end{figure}

A popular trading policy is based on \ac{bb}, which is a simple and fundamental technique employed in a variety of trading schemes~\cite{bollinger1992using}.  
\ac{bb} consists of 3 bands plotted around the asset's price -- upper, middle, and lower -- as illustrated in Fig.~\ref{fig:bb_figure}. The middle band is a simple \ac{ma}, whose window size varies per application (in Fig.~\ref{fig:bb_figure} we used a window of 20 samples). The top and bottom bands are plotted  around the middle band where the distance can be based on the \ac{std} of the \ac{ma}. These are typically set at $\pm$ 1 \ac{std} around the \ac{ma}, though the setting may vary depending on the application. Alternatively, one may use confidence intervals for forming such bands.

Using these bands, one can build a trading strategy. A natural approach to do so is applicable when $\gscal{d}_\tau$ is a stationary price series. In this case, one can construct a financial tracker using the empirical $z$-score, i.e.,
\begin{equation}
 \gscal{z}_t = \varphi\left(\{\gscal{d}_\tau\}_{\tau\leq t}\right) = \frac{\gscal{d}_t - \mu_t}{\sigma_t},
 \label{eqn:Zscore}
\end{equation}
where $\mu_t$ and $\sigma_t$ are the empirical first and second order
moments of $\gscal{d}_t$, respectively, estimated from $\{\gscal{d}_\tau\}_{\tau\leq t}$. 

The \ac{bb} policy is obtained by examining in which band $\gscal{z}_t$ lies. In particular, if an open position is not currently being held, a short position is taken if the asset is being overbought, i.e. $\gscal{z}_t > 1$,  and a long position if its being oversold , i.e. $\gscal{z}_t < -1$. To formulate this mathematically, we say that an open position is held at time $t$ if the last open position time denoted
\begin{equation}
\label{eqn:LastOpen}
    \tau_{{\rm op},t} \triangleq \max_{\tau < t: \gscal{op}_\tau = \pm 1} \tau,
\end{equation}
is not smaller than the last close position time
\begin{equation}
\label{eqn:LastClose}
    \tau_{{\rm cp},t} \triangleq \max_{\tau \leq t: \gscal{cp}_\tau =  1} \tau. 
\end{equation}
The open position policy is thus determined as
\begin{equation}
\label{eqn:OpenPolicy}
    \gscal{op}_t = \left(\unitstep{-1 - \gscal{z}_t} -\unitstep{\gscal{z}_t -1}\right) \cdot \unitstep{\tau_{{\rm cp},t} -  \tau_{{\rm op},t}}.
\end{equation}

The reward in \eqref{eq:reward} is formulated for each {\em active position}, and not for each time instance. In some settings, e.g., when designing trading strategies using \ac{rl} \cite{ayadireinforcement,brim2020deep,kim2022hybrid}, one is often interested in obtaining {\em instantaneous rewards}. This  achieved by   closing a position after a single time step (though it can then re-opened and treated as a new active position, yielding an addition transaction cost, i.e., {\em friction}~\cite{krauss2017statistical}, which we omit for simplicity). Such an operation results in
\begin{equation}
\label{eqn:ClosePolicyInst}
   \gscal{cp}_t = \unitstep{|\gscal{op}_{t-1}|}.
\end{equation}
 Alternatively, one can determine the close position based on the indicator, allowing   a {\em cumulative reward} where a position can be held over multiple time steps. In this case, the closing of a position is a function of the indicator $\gscal{z}_t$. For instance,  one can decide to close a currently open position if  $\gscal{z}_t$ has crossed the middle or opposite band \cite{kannan2010financial,stubinger2017statistical,chan2013algorithmic}, i.e., 
 \begin{equation}
\label{eqn:ClosePolicyCum}
   \gscal{cp}_t =    \unitstep{-\gscal{z}_{t-1} \cdot \gscal{z}_{t}} \cdot (1 - \unitstep{\tau_{{\rm cp},t} -  \tau_{{\rm op},t}}).
\end{equation}

\begin{figure}
  \centering
  \includegraphics[width=0.5\textwidth]{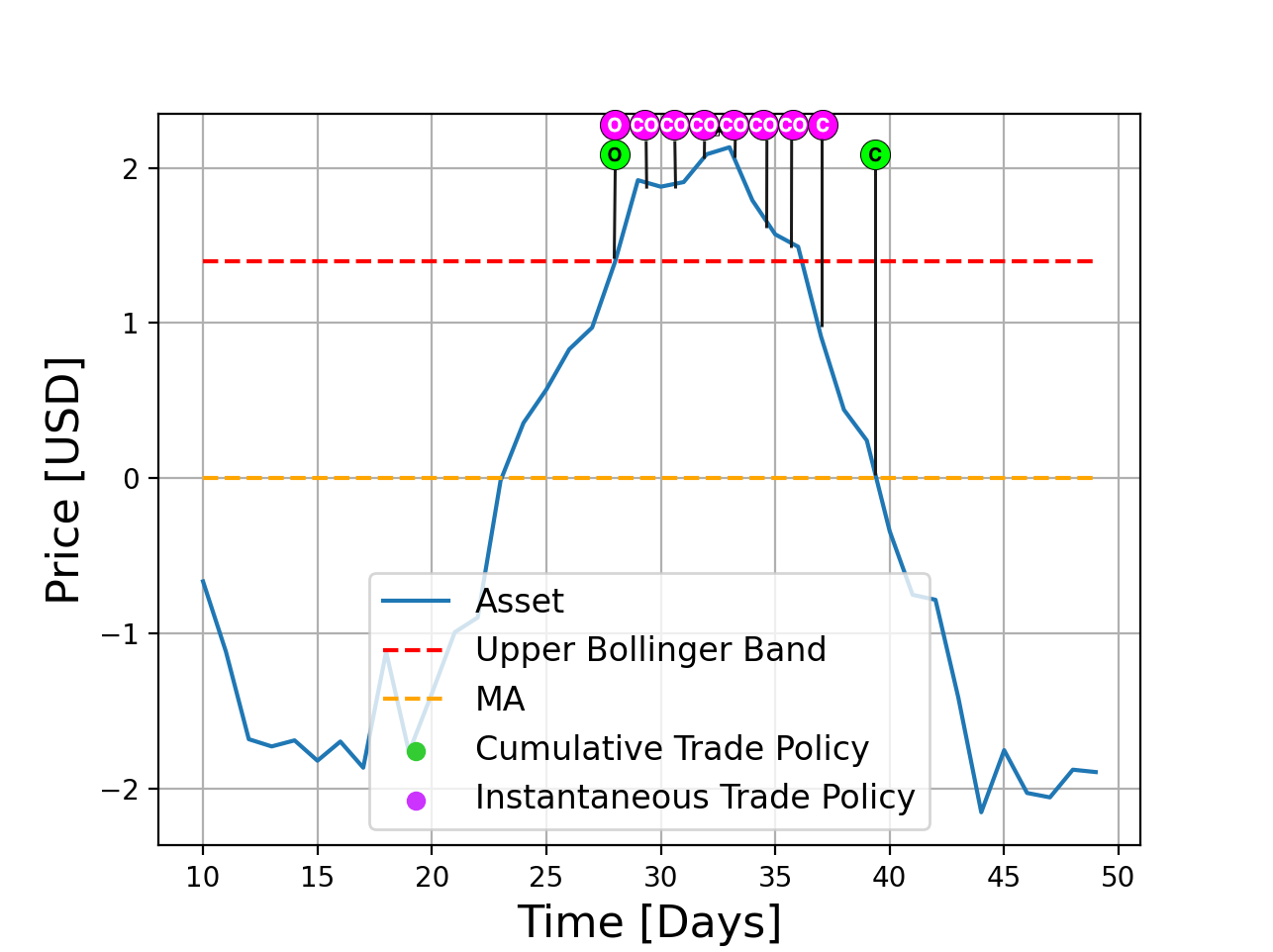}
  \caption{Comparison between the trading actions of the cumulative trading policy (green) and an instantaneous trading policy along with the \ac{ma} and upper \ac{bb} limit. Here {\em C} closes a position, {\em O} opens a position, and {\em CO} closes and  opens a position on the same day.}
  \label{fig:policy_comparison}
\end{figure}

A comparison between the cumulative and instantaneous approaches is schematically illustrated in Fig.~\ref{fig:policy_comparison}, where only  the middle and upper Bollinger Bands are drawn for clarity. As depicted in  Fig.~\ref{fig:policy_comparison},  the instantaneous trading policy  effectively makes  profit based on the difference between the asset price when it crosses the upper band for the second time and the price of the asset on the following day. On the other hand, the profit made by the cumulative trading policy is the the price difference between the upper band and the \ac{ma} of the asset.

There are several advantages in using \ac{bb} for determining trading policies, that go beyond the profit-loss reward, and stem from the simplicity and interpretability of the scheme. First, the bands adjust to the volatility of the asset's price, enabling to visually represent it. In addition, \ac{bb} can be used to determine price reversal signals - when the price of the asset goes above the upper band or below the lower band it can indicate that the asset is being overbought or oversold, respectively, and therefore suggest that a trend reversal is a likely to occur. Moreover, \ac{bb} can identify trends - if the price is constantly above the middle band it can suggest a uptrend or similarly a downtrend if its below the middle band \cite{kannan2010financial}.

%
\subsection{Problem Formulation}\label{ssec:problem} 
We consider the design of a trading strategy for the task of {\em pairs trading}. Here, the financial information accumulated at each time instance $t$ corresponds to a pair of assets denoted $\alpha_t$ and $\beta_t$. There is no underlying assumption on stationarity on the considered assets.
Our goal is to jointly design a financial  tracker $\varphi$ along with a policy mapping $\pi$ that maximizes the  expected \ac{pnl}. The \ac{pnl} is defined as the sum of all rewards accumulated up to time $t$, i.e.,
\begin{equation}
 \label{eq:PNL}
 {\rm PNL}^{\varphi,\pi}_t=\sum_{i: t_{{\rm cp}_i} \leq \tau_{{\rm cp},t}}\gscal{r}_{i}^{\varphi,\pi}.  
 \end{equation} 
Accordingly, the trading strategy design problem is formulated as identifying
\begin{equation}
\label{eqn:Problem}
  \varphi^\ast, \pi^\ast= \arg\max_{\varphi, \pi}\mathbb{E}\set{{\rm PNL}^{\varphi,\pi}_t}.  
\end{equation}


For design purposes, we are given access to a data set comprised of past financial information measurements corresponding to $n_t$ past time indices. This dataset is given by
\begin{equation}
\label{eqn:Dataset}
    \mySet{D}=\{\alpha_\tau, \beta_\tau\}_{\tau=-n_t}^{-1}.
\end{equation}
In order to leverage established \ac{bb} techniques, which inherently rely on underlying stationarity, we exploit pairs trading modeling, where we extend the notions of \ac{ci} and \ac{pci} proposed in \cite{puspaningrum2012pairs,clegg2018pairs}, as detailed in the following Section~\ref{sec:State}. This enables a hybrid model-based/data-driven design, as stated in Section~\ref{sec:Nerual_Aug_TP}.

\section{Pairs Trading as a State Space Model}
\label{sec:State}
The price of a single asset is rarely described as stationary and is typically a chaotic stochastic process. However, tracking some inherit  statistical relationship between two assets is often much easier  than tracking each asset individually, and consequently can be used for investment decision. This is the reasoning behind pairs trading, and its general form which consists of more than two assets~\cite[Ch. 10]{feng2016signal}. 

There are several frameworks for capturing statistical relationships in pairs trading. These include monitoring distances~\cite{bajalan1999pair}, adopting a \ac{ci} model~\cite{puspaningrum2012pairs}, as well as viewing pairs trading as a stochastic control  setup~\cite{mudchanatongsuk2008optimal}. A large volume of of research is dedicated to comparing these statistical models and their usefulness for pairs trading~\cite{krauss2017statistical}. In this work, we adopt the \ac{pci} approach~\cite{clegg2018pairs}, which extends the \ac{ci} model. The \ac{pci} model is  suitable for tackling~\eqref{eqn:Problem} due to its ability to capture both long-lasting and transient company shocks and their effect on the asset price, and its flexibility in describing temporal correlations in the pairwise relationship~\cite{clegg2018pairs}. 

To describe this model, we first review the notion of \ac{ci} and how it gives rise to a \ac{ss} model for pairs trading  in Subsection~\ref{ssec:CI_review}. Then, in Subsection~\ref{ssec:PCI_MODELs} we review the extended \ac{pci} model and  present our \ac{ss} representation for pairs trading that follows from the \ac{pci} model. The proposed extended \ac{ss} model is used to derive  \ac{kbpt} in Section~\ref{sec:Nerual_Aug_TP}.


\subsection{Co-Integration  in Pairs Trading}\label{ssec:CI_review}
We next review the notion of \ac{ci} how it is specialized for  pairs trading. 

\subsubsection{Co-Integration}
To define \ac{ci}, we first recall the definition of integration order (see, e.g., \cite[Ch. 2.6]{feng2016signal}):
\begin{definition}[Integration Order]
    \label{def:integration}
A time series  $\gscal{x}_t$ is said to be {\em integrated with order $p$}, and written as $\gscal{x}_t \sim \gscal{I}(p)$, if  there exists a $(p+1)\times 1$  vector $\myVec{w}$ such that $\myVec{w}^T\myVec{x}_t$ is a stationary time series, with  $\myVec{x}_t \triangleq [\gscal{x}_t,\gscal{x}_{t-1},...,\gscal{x}_{t-p}]$. 
\end{definition}

The definition of \ac{ci} generalizes  Definition~\ref{def:integration} by considering multivariate time sequences as follows:
\begin{definition}[Co-Integration]
    \label{def:Cointegration}
A $(p+1)\times 1$ multivariate time sequence $\bar{\myVec{x}}_t = [\gscal{x}_{t}^{(1)},\gscal{x}_{t}^{(2)},...,\gscal{x}_{t}^{(p+1)}]$ is said to obey a {\em \ac{ci} model of order $p$}, denoted $\bar{\myVec{x}}_t \sim \gscal{CI}(p)$,  if there exists a $(p+1)\times 1$  vector $\myVec{w}$ such that $\myVec{w}^T\myVec{x}_t$ is stationary. 
\end{definition}
Definition~\ref{def:Cointegration} clearly specializes  Definition~\ref{def:integration}, as for a time series $\gscal{x}_t$ with integration order $p$, one can equivalently define the co-integrated multivariate time sequence $\bar{\myVec{x}}_t$ by having its $\tau$th entry, denoted $\gscal{x}_{t}^{(\tau)}$, set to $\gscal{x}_{t}^{(\tau)} = \gscal{x}_{t-\tau+1}$ for every $\tau \in \{1,\ldots,p+1\}$.

\subsubsection{Co-Integration in Pairs Trading}\label{ssec:CI_model_review} 
\ac{ci}  is often used to model pairs trading. In this framework, the two assets $\alpha_t$ and $\beta_t$ are assumed to obey a $\gscal{CI}(1)$ model (see Def.~\ref{def:Cointegration}). In particular, we define the {\em spread} time sequence $\gscal{s}_t$ via
\begin{equation}
\label{eqn:spreadCI}
 \gscal{s}_t = \beta_t - \gscal{h}\cdot\alpha_t - \mu,
\end{equation}
and look for $h$ and $\mu$ such that $\gscal{s}_t$ is a zero-mean stationary time series. We refer to $h$ as the {\em hedge ratio} and $\mu$  as the {\em equilibrium value}. 
These parameters are typically estimated by rewriting~\eqref{eqn:spreadCI} as
\begin{equation}\label{eq:CI_model_constant_coeff}
 \beta_t = \gscal{s}_t + \gscal{h}\cdot\alpha_t + \mu,
\end{equation}
from which $ \gscal{h}$ and  $ \mu$ can  be recovered from a set of pairs as in~\eqref{eqn:Dataset} via, e.g., least-squares~\cite{vidyamurthy2004pairs}.
Once $ \gscal{h}$ and  $ \mu$ are estimated, one can apply statistical tests to check for stationarity, e.g.,   Dicky-Fuller test~\cite{dickey1979distribution} or Johansen test~\cite{johansen1991estimation}.

\subsubsection{Co-Integration SS Model}\label{ssec:CI_SS_Model} 
An inherent limitation with the above form of \ac{ci} modelling of pairs trading follows from the fact that  $ \gscal{h}$ and  $ \mu$ are assumed to be static. In practice, they might drift, and the resulting model may not be reliable to serve as a basis for designing trading policies. However, the \ac{ci} framework in~\eqref{eq:CI_model_constant_coeff} can be used to form a \ac{ss} model that supports tracking of the hedge ratio and equilibrium values using, e.g., the \ac{kf}. 

The \ac{ss} model for pairs trading with the \ac{ci} framework assumes that $ \gscal{h}$ and  $ \mu$ are first-order Markov processes  
\begin{subequations}\label{eq:CI_SS_Model}
 \begin{align}
  &\gscal{h}_t = \gscal{h}_{t-1} + \epsilon_t^h,\\
  &\mu_t = \mu_{t-1} + \epsilon_t^\mu,
 \end{align}
where $\epsilon_t^h$ and $\epsilon_t^\mu$ follow an i.i.d. Gaussian distribution. Accordingly, the relationship between the assets pairs in \eqref{eq:CI_model_constant_coeff} is replaced with
\begin{align}
\label{eq:CI_model_constant_coeff2}
&\beta_t = \gscal{s}_t + \gscal{h}_t\cdot\alpha_t + \mu_t.
\end{align}
\end{subequations}

The representation in \eqref{eq:CI_SS_Model} gives rise to a \ac{ss} model  as follows. The latent state  vector is $\gvec{x}_t=[\gscal{h}_t,\mu_t]^T$, the observation is set to be $\gscal{y}_t=\gscal{\beta}_t$, and  the underlying dynamics of the \ac{ss} model are 
\begin{subequations}\label{eq:SS_model}
\begin{align}
\gvec{x}_{t}= \gvec{F}_t\cdot\gvec{x}_{t-1}+\gvec{e}_t = \begin{bmatrix}
    1 & 0 \\
    0 & 1 \\
    \end{bmatrix}\gvec{x}_{t-1} + \gvec{e}_t,\label{eq:SS_model_ss_vector}\\
\gscal{y}_{t}=
\gvec{g}_t^T\cdot\gvec{x}_{t}+\gvec{v}_{t}=\begin{bmatrix}
    \alpha_t &
    1 \\
    \end{bmatrix}\gvec{x}_t + \gscal{v}_{t}.\label{eq:SS_model_obs}
\end{align}
\end{subequations} 
In \eqref{eq:SS_model_obs}, $\gscal{v}_{t}$ is modeled as white Gaussian observation noise, and the measurement noise is given by $\myVec{e}_t\triangleq[\epsilon_t^h,\epsilon_t^\mu]^T$. 

Formulating the evolution of the hedge ratio and equilibrium value as a linear Gaussian \ac{ss} via \eqref{eq:SS_model} allows the tracking of these parameters using the \ac{kf}. 
The \ac{kf} is comprised of  prediction  and  update stages, respectively given as
\begin{subequations}\label{eq:KF_eqs}
\begin{align}\label{eq:KFPredict}
&\hat{\gvec{x}}_{t|t-1}= \gvec{F}_t\cdot\hat{\gvec{x}}_{t-1},
\hspace{0.2cm}
&\hat{\gscal{y}}_{t|t-1}=\gvec{g}_t^T\cdot \hat{\gvec{x}}_{t|t-1},\\
&\hat{\gvec{x}}_{t}=
\Kgain_t \cdot \Delta\gscal{y}_t + \hat{\gvec{x}}_{t|t-1}, 
\hspace{0.2cm}
&\Delta\gscal{y}_t = 
\gscal{y}_t - \hat{\gscal{y}}_{t|t-1}.\label{eq:KFUpdate}
\end{align}
\end{subequations}
In \eqref{eq:KFUpdate}, $\Kgain_t$ is the \ac{kg} computed from the second-order moments, and $\Delta\gscal{y}_t$ is the innovation term. 

By \eqref{eq:CI_model_constant_coeff2}, $\Delta\gscal{y}_t$
can be written as
\begin{align}\label{eq:SSinnovation1}
\Delta\gscal{y}_t &=  \gscal{y}_t -  \hat{\gscal{h}}_{t|t-1} \gscal{\alpha}_t - \hat{\gscal{\mu}}_{t|t-1},
\end{align}
and thus it describes the difference between the posterior and the prior.
Using the fact that $\gscal{y}_t = \gscal{\beta}_t$  and substituting the time varying relationship between $\alpha_t$ and $\beta_t$ from  \eqref{eq:SSinnovation1}, we obtain that
\begin{align}\label{eq:SSinnovation2}
\Delta\gscal{y}_t &=   \gscal{s}_t + \left((\gscal{h}_t - \hat{\gscal{h}}_{t|t-1}) + (\gscal{\mu}_t -\hat{\gscal{\mu}}_{t|t-1})\right). 
\end{align}
From  \eqref{eq:SSinnovation2} it follows that $\Delta\gscal{y}_t$ is an estimate of the value of the spread corrupted by an additive term that represents the estimation error at time $t$. Thus, $\Delta\gscal{y}_t$ can be used as an equivalent pair-wise asset, from which one can extract a useful indicator for \ac{bb} as in \eqref{eqn:Zscore},  i.e., 
%
\begin{equation}
\label{eq:CI_MODEL_INDICATOR}
\gscal{z}_t = \frac{\Delta \gscal{y}_t}{\gscal{\sigma}^y_{t}} .
\end{equation}
Here, $\gscal{\sigma}^y_{t}$ is the \ac{std} of $\Delta\gscal{y}_t$ (which is also tracked by the \ac{kf} for computing the \ac{kg}).

\subsection{Partial Co-Integration in Pairs Trading}\label{ssec:PCI_MODELs}
While the \ac{ci} model for pairs trading gives rise to a simple \ac{kf}-aided trading policy, it is limited in the sense that it does not incorporate any temporal statistical model on the spread time sequence, which is at the core of the financial indicator. Such a temporal statistical model is incorporated by extending the notion of \ac{ci} into \ac{pci}~\cite{clegg2018pairs}. In the following, we first recall the \ac{pci} definition, after which we review its associated \ac{ss} model proposed in \cite{clegg2018pairs}, and then further extend it to propose the \ac{ss} model for pairs trading used in our derivation in Section~\ref{sec:Nerual_Aug_TP}. 

\subsubsection{Partial Co-Integration}\label{ssec:PCI_review} 
An extended definition of \ac{ci} is that of \ac{pci}~\cite{clegg2018pairs}:
\begin{definition}[Partial Co-Integration]
    \label{def:PartialCointegration}
A  multivariate time sequence $\bar{\myVec{x}}_t$ is said to obey a {\em \ac{pci} model of order $p,q$}, denoted $\bar{\myVec{x}}_t \sim \gscal{PCI}(p,q)$,  if there exists a   vector $\myVec{w}$ such that the time series $\myVec{w}^T\myVec{x}_t$ is combination of scalar time sequences that are integrated in time with orders $q$ and $p-q$. Namely, it can be decomposed into $\myVec{w}^T\myVec{x}_t=\mu_t+\gscal{s}_t$ where $\mu_t\sim \gscal{I}({q})$ and $\gscal{s}_t\sim \gscal{I}(\gscal{p}-{q})$.
\end{definition}

Note that any \ac{ci} multivariate time sequence  (Definition~\ref{def:Cointegration}) is also \ac{pci}  and is obtained when $q= p = 0$, which results in $\myVec{w}^T\myVec{x}_t$ being a stationary time series.

\subsubsection{Partial Co-Integration in Pairs Trading}\label{ssec:PCI_model_review} As suggested in ~\cite{clegg2018pairs},  the \ac{pci} model can be used to generalize the \ac{ci} model for pairs trading. In this case, it is  assumed that $[\beta_t, \alpha_t]~\sim~\gscal{PCI}(1,1)$, i.e.,
\begin{subequations}\label{eq:PCI_Model}
 \begin{align}
 &\beta_t  =\gscal{s}_t  + \gscal{h}\cdot\alpha_t +\mu_t,
 \end{align}
 where 
  \begin{align}
  \label{eqn:spreadEvolve}
 &\gscal{s_t} = \rho\gscal{s_{t-1}} + \epsilon^s_t, \\
 &\mu_t = \mu_{t-1} + \epsilon_t^{\mu}.
 \end{align}
\end{subequations}
The \ac{pci} model for pairs trading preserves the notions of {spread} ($\gscal{s}_t$) and hedge ratio ($\gscal{h}$) - where in this model , the hedge ratio is constant unlike in the \ac{ci} model which it was modeled as time varying. It introduces the parameter $\rho \in (-1,1)$ as an autoregression coefficient, while $\epsilon_t^\mu$, $\epsilon_t^s$ are stationary signals, often modeled as mutually independent Gaussian white noises~\cite{clegg2018pairs}.

The \ac{pci} framework in \eqref{eq:PCI_Model} introduces two main degrees of freedom compared with the \ac{ci} model in \eqref{eq:CI_model_constant_coeff}:  
$(i)$ First, the \ac{ci} model does not explicitly capture the temporal correlation in the spread time sequence $\gscal{s}_t$, where  the \ac{pci} model incorporates such correlation via its autoregressive modelling and the parameter er  $\rho$;  
$(ii)$  \ac{pci} captures  temporal variations in the  equilibrium value $\mu_t$, by modelling it  as a random walk. 

\subsubsection{Partial Co-Integration SS Model}\label{ssec:PCI_SS_Model} 
The \ac{ss} model proposed in \cite{clegg2018pairs} for using the \ac{pci} representation in \eqref{eq:PCI_Model} for trading policies sets the latent state vector to be $\gvec{x}_t = [\beta_t,\gscal{s}_t,\mu_t]^T$, the observation to be $\gvec{y}_t=[\gscal{\alpha_t},\gscal{\beta_t}]^T$, and the underlying dynamics of the \ac{ss} model are 
\begin{subequations}
\label{eq:PCIoldss}
    \begin{align}
        \gvec{y}_t &= \gvec{H}_t\gvec{x}_t =
        \begin{bmatrix}
        h & 1 & 1 \\
        1 & 0 & 0 \\
        \end{bmatrix}\gvec{x}_t, \\
        \gvec{x}_t &= \gvec{F}_t\gvec{x}_{t-1} + \gvec{q}_t \notag \\ 
        &=
        \begin{bmatrix}
        1 & 0 & 0 \\
        0 & \rho & 0 \\
        0 & 0 & 1 \\
        \end{bmatrix} \gvec{x}_{t-1} +
        \begin{bmatrix}
        \epsilon_t^\beta \\
        \epsilon_t^s \\
        \epsilon_t^\mu \\
        \end{bmatrix}.
    \end{align}
\end{subequations}
In \eqref{eq:PCIoldss}, $\gvec{q}_t$ is a multivariate white Gaussian noise. 

Note that the \ac{ss} model in \eqref{eq:PCIoldss}, which is the one suggested in~\cite{clegg2018pairs}, differs from the \ac{ci}-based \ac{ss} model of \eqref{eq:SS_model} not only in its usage of the \ac{pci} model, but also in the fact that the spread is considered in the state vector, while the hedge ratio is assumed to be static. Accordingly, one can track the state using the \ac{kf} (given in \eqref{eq:KFPredict}-\eqref{eq:KFUpdate}) respectively,  and use the tracked spread to form  the indicator used for \ac{bb} trading as
\begin{equation}\label{eq:PCI_Indicator}
    \gscal{z}_t = \frac{\hat{\gscal{s}}_t}{\sigma_t^s}
\end{equation}
where $\sigma_t^s$ is the \ac{std} of the spread.

\subsubsection{Proposed Partial Co-Integrated \ac{ss} Model} \label{ssec:Our_PCI_SS_Model}
While the \ac{pci} model in \eqref{eq:PCIoldss} offers additional degrees of freedom in capturing temporal dependencies in the spread, it is associated with two main drawbacks:
$(i)$ the hedge ratio is modeled as being static, which, as stated before might not always be the case; 
and $(ii)$ the model  assumes that all the noises are Gaussian, i.i.d. and mutually independent. In financial data, this is rarely the case \cite[Ch. 3]{feng2016signal}.  
Accordingly, we propose an alternative \ac{ss} model for pairs trading which shares the improved temporal degrees of freedom of the \ac{pci} model, while accounting for the dynamic nature of the hedge ratio, without imposing a specific distribution on the stochasticity. 

Our proposed \ac{ss} Model is similar to the \ac{ci} \ac{ss} model in \eqref{eq:SS_model}, while we also add the assumed autoregressive behavior of the spread in \eqref{eqn:spreadEvolve}. 
Specifically,  our latent space vector is  $\gvec{x}_t =~[\gscal{h}_t,\gscal{\mu}_t,\gscal{s}_t]^T$, the observation is $\gscal{y}_t=\gscal{\beta_t}$, and the dynamics of the systems are modeled as
\begin{subequations}
\label{eqn:OurSSModel}
    \begin{align}
        \gvec{x}_t &= \gvec{F}_t\gvec{x}_{t-1} + \gvec{e}_t=\begin{bmatrix}
            1 & 0 & 0\\
            0 & 1 & 0\\
            0 & 0 & \rho \\
        \end{bmatrix}\gvec{x}_{t-1} + \gvec{e}_t, \\
        \gscal{y}_t &= \gvec{g}_t^T\gvec{x}_t + \gscal{v}_t = \begin{bmatrix}
            \alpha_t &
            1 &
            1 \\      
        \end{bmatrix}\gvec{x}_t + \gscal{v}_t.
    \end{align}
\end{subequations}
It is  emphasized that we do not impose a specific distribution  on the noise vectors $\gvec{e}_t$ and $\gscal{v}_t$, unlike in \eqref{eq:SS_model} and \eqref{eq:PCIoldss}, where Gaussian noise are assumed.  


\section{Neural Augmented Trading Policy}
\label{sec:Nerual_Aug_TP}

We are interesting in exploiting the \ac{ss} model representation in \eqref{eqn:OurSSModel} for formulating a trading policy following established approaches combining \ac{kf} and \ac{bb} policies, e.g., \cite{clegg2018pairs}. However, the application of such methodologies gives rise to the following challenges:
\begin{enumerate}[label={\em C\arabic*}] 
    \item \label{itm:model} {\em Modelling Stochasticity:} \ac{kf} inherently assumes Gaussian noise.  In financial data, this is unlikely to hold, and the noise tends to follow a heavy tailed distribution~\cite{bradley2003financial}.
    \item \label{itm:mismatch} {\em Modelling Misamtches:} the models for $\gvec{F}_t$ and $\gvec{g}_t$ in \eqref{eqn:OurSSModel}, which follow from the \ac{pci}, are  an approximation of the complex real-world dynamics governing pairs trading. Therefore, a model mismatch is inevitable.
    \item \label{itm:pnl} {\em State Tracking:} Tracking based on \ac{kf} is inherently designed to provide an accurate estimate of its state vector $\myVec{x}_t$. Here, $\myVec{x}_t$ is comprised of  model parameters that do not necessarily represent a physical quantity, but are rather intermediate variables used for trading policy. What one is really interested in is obtaining a representation that is most useful for pairs trading in the \ac{pnl} sense. However, the conventional  \ac{kf} is by design tailored to optimize tracking the state, which are here assumed model parameters, though this might be sub-optimal for maximizing the \ac{pnl}.
\end{enumerate}

To enable pairs trading under \ref{itm:model}-\ref{itm:pnl}, we propose a hybrid model-based/data-driven algorithm based on neural augmentation of \ac{kf}-\ac{bb} pairs trading. We present the proposed \ac{kbpt} and its training procedure in in Subsections~\ref{ssec:kbpt}-\ref{ssec:training}, respectively, and a discussion is provided in Subsection~\ref{ssec:discussion}

\subsection{\acl{kbpt}}
\label{ssec:kbpt}
\subsubsection{High Level Design}
We design our pairs trading scheme to preserve the interpretable and reliable operation of the combination of \ac{kf} with \ac{bb} trading under our proposed \ac{ss} model in \eqref{eqn:OurSSModel}. To cope with challenges \ref{itm:model}-\ref{itm:mismatch}, we augment the operation of the \ac{kf}, and specifically its computation of the \ac{kg}, with a dedicated \ac{rnn}, following the \acl{kn} algorithm proposed in \cite{revach2022kalmannet}. \acl{kn} systematically uses the available model parameters  $\gvec{F}_t$ and $\gvec{g}_t$ while using an \ac{rnn} to bypass the need to track the statistical moments of the dynamics, allowing it to overcome  \ref{itm:model}-\ref{itm:mismatch}. To cope with \ref{itm:pnl}, we leverage the trainability of \acl{kn} to adapt the overall policy based on the \ac{pnl} objective, i.e., to provide a financial indicator that is most useful for trading, rather than arising from the most accurate description of the state. This is not always possible with fully model-based methods, which inherently separate tracking from policy making, and thus the tracking method is invariant the main objective, i.e., maximizing the \ac{pnl}.

\begin{figure}
  \centering
  \includegraphics[width=0.5\textwidth]{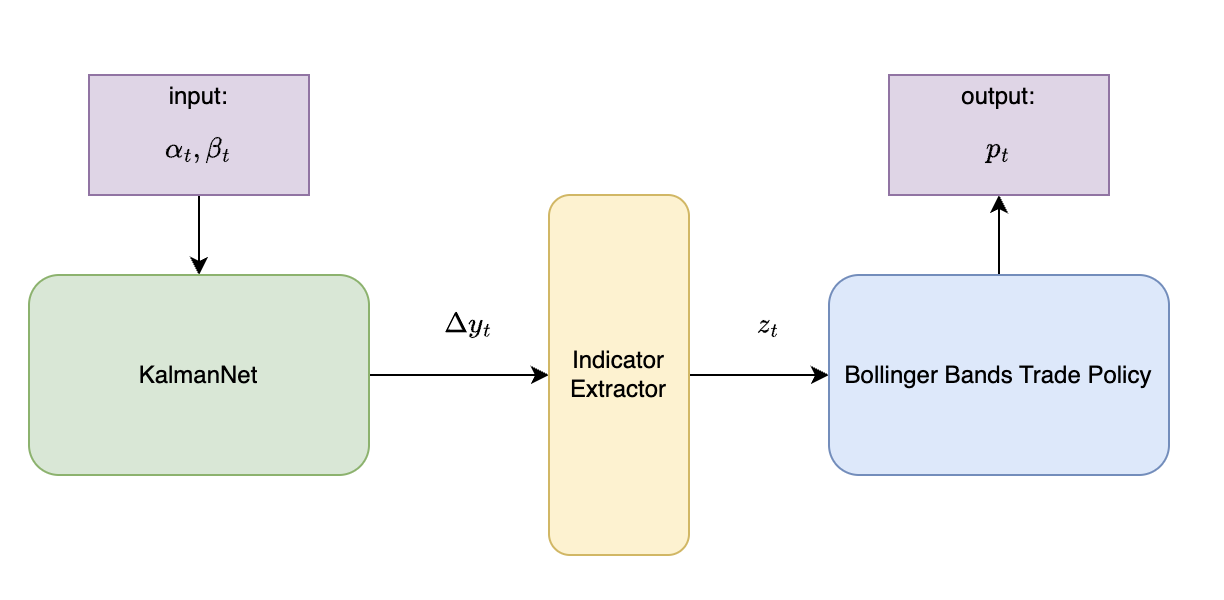}
  \caption{\ac{kbpt} pipeline.}
  \label{fig:PT_pipeline}
\end{figure}

\subsubsection{\ac{pci} \ac{ss} Tracking with KalmanNet}
Our pairs trading scheme thus consists of three components, depicted in Fig.~\ref{fig:PT_pipeline}. The first component is \acl{knet}, which takes as input the two assets used for pairs trading, tracks the state vector under the \ac{ss} model of \eqref{eqn:OurSSModel}, and outputs the innovation $\Delta \gscal{y}_t$. 

{\acl{kn} augments  the \acl{mb} \ac{kf} with an \ac{rnn} to cope with model mismatch and non-linearities. The \ac{rnn}, whose trainable parameters are denoted by $\myVec{\theta}$ is used to output the \ac{kg}, denoted here $\Kgain_t\brackets{\myVec{\theta}}$. Since the computation of the \ac{kg} in the \ac{kf} encapsulates the dependency on the distribution of the noise signals, replacing it with a trainable architecture bypasses the need to impose such a model and enables us to model complex, non-linear noises due to the underlying architecture of the \ac{rnn}, while preserving the operation of the \ac{kf}. Accordingly, the prediction and update stages of \acl{kn} are given by
\begin{subequations}
\label{eqn:KNop}
\begin{align}\label{eq:KNPredict}
&\hat{\gvec{x}}_{t|t-1}= \gvec{F}_t\cdot\hat{\gvec{x}}_{t-1},
\hspace{0.2cm}
&\hat{\gscal{y}}_{t|t-1}=\gvec{g}_t^T\cdot \hat{\gvec{x}}_{t|t-1},\\
&\hat{\gvec{x}}_{t}=
 \Kgain_t\brackets{\myVec{\theta}}\cdot \Delta\gscal{y}_t + \hat{\gvec{x}}_{t|t-1}, 
\hspace{0.2cm}
&\Delta\gscal{y}_t = 
\gscal{y}_t - \hat{\gscal{y}}_{t|t-1}.\label{eq:KNUpdated}
\end{align}
\end{subequations}
%
The sole  difference between \eqref{eqn:KNop} and \eqref{eq:KF_eqs} is that here in the update stage in \eqref{eq:KNUpdated} utilizes the \ac{kg} computed by the \ac{rnn}, instead of computing it from knowledge of the underlying statistics.
\acl{kn} leverages the \ac{ss} representation of pairs trading and maintains the low latency and complexity of \ac{kf}-aided tracking while coping with its inevitable mismatches and approximation errors. The detailed architecture of the \ac{rnn} is given in \cite{revach2022kalmannet}. 

\subsubsection{\ac{bb}-Based Trading}
The innovation produced by \acl{kn} is  passed  to the indicator extractor which outputs its $Z$ score, denoted  $\gscal{z}_t$, via
\begin{equation}\label{eq:OurPCIIndicator}
    \gscal{z}_t = \frac{\Delta{\gscal{y }}_t}{\hat\sigma_t}.
\end{equation} 
 In \eqref{eq:OurPCIIndicator}, $\hat\sigma_t$ is the empirical \ac{std} of the innovation process calculated using a rolling window. In our experimental study reported in Section~\ref{sec:emp_eval} we used a window size of $80$ samples based on empirical trials. The architecture of  \acl{knet} with the indicator extractor is depicted in Fig.~\ref{fig:KNET_indicator_diagram}.

\begin{figure*}
  \centering
  \includegraphics[width=0.8\textwidth]{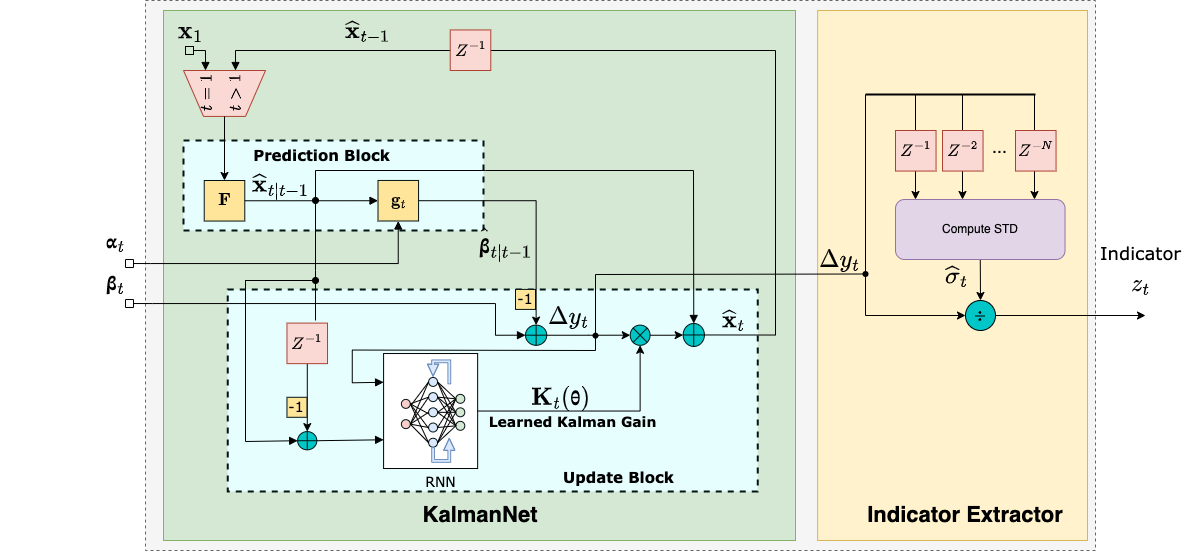}
  \caption{\acl{knet} with indicator extraction for the \ac{bb} trade policy.}
  \label{fig:KNET_indicator_diagram}
\end{figure*}

 The final component of our trading policy executes the \ac{bb} trade policy on $\gscal{z}_t$, Its output is the position to take at time $\gscal{t}$ denoted as $\gscal{p}_t$ based on \eqref{eqn:OpenPolicy} for opening and  \eqref{eqn:ClosePolicyCum} for closing. The resulting overall trading policy is summarized as Algorithm~\ref{alg:Trading}.
%
%

%
\begin{algorithm}
\caption{Neural Augmented \ac{kbpt} at Time $t$}
\label{alg:Trading} 
\SetKwInOut{Initialization}{Init}
\Initialization{\ac{rnn} parameters $\myVec{\theta}$}
\SetKwInOut{Input}{Input} 
\Input{Previous estimate $\hat{\gvec{x}}_{t-1}=[\hat{\gscal{h}}_{t-1},\hat{\gscal{\mu}}_{t-1},\hat{\gscal{s}}_{t-1}]^\top$;\\ 
Assets $\beta_t, \alpha_t$.}  
\nonl\texttt{\acl{kn}:}\\
Predict $\hat{\gscal{y}}_{t|t-1}$ via \eqref{eq:KNPredict}\;
Update estimate  and compute $\Delta \gscal{y}_t$ via   \eqref{eq:KNUpdated}\;
\nonl\texttt{Indicator Extraction:}\\
Calculate  feature $\gscal{z}_t$ via \eqref{eq:OurPCIIndicator}\;
\nonl\texttt{Bollinger Bands:}\\
Compute $\gscal{p}_t$ via \eqref{eqn:OpenPolicy} and \eqref{eqn:ClosePolicyCum} with $\gscal{z}_t$ as argument\;
\end{algorithm}

\subsubsection{Reward}
The \ac{bb} policy is stated in Subsection~\ref{ssec:Bollinger} for a single asset. Here, since we consider pairs trading,  an open position taken for the $i$th transaction at time $\topen$ implies  open positions on $\alpha,\beta$ in the sum of $1\$$. Its division between the assets is based on the current hedge ratio, i.e., $\hat{\gscal{h}}_{\topen}$. Similarly,  when the position is closed at time $\tclose$, we close the open positions on $\alpha,\beta$ according to the hedge $\hat{\gscal{h}}_{\tclose}$. Accordingly, the reward of each asset in such a transaction is the sum of the rewards in $\alpha$ and in $\beta$, i.e., 
\begin{equation} \label{eq:total_reward_per_transaction}
     r^{\varphi,\pi}_i =  r^{\varphi,\pi,\beta}_i  + r^{\varphi,\pi,\alpha}_i.
\end{equation}
To formulate the individual reward terms in \eqref{eq:total_reward_per_transaction}, recall that ${\rm op}_{\topen}$ is the position taken on the \textit{spread}. 
Consequently, to set the position taken with respect to $\beta$ and $\alpha$ individually it should be multiplied by $\zeta_{\topen} \triangleq {\rm sign}(\beta_{\topen} - \hat{\gscal{h}}_{\topen}\alpha_{\topen})$ and $-\zeta_{\topen}$, respectively, to determine what position is taken on each \textit{individual asset} given a position on the spread. The resulting reward terms are thus
\begin{subequations}\label{eq:assets_return}
 \begin{align}
 r^{\varphi,\pi,\beta}_i =& \left(\frac{\beta_{\tclose}}{1+|\hat{\gscal{h}}_{\tclose}|} - \frac{\beta_{\topen}}{1+|\hat{\gscal{h}}_{\topen}|}\right) \cdot {\rm op}_{\tclose} \cdot 
 \zeta_{\topen}, \\
 r^{\varphi,\pi,\alpha}_i  = &\left(\frac{|\hat{\gscal{h}}_{\topen}|\alpha_{\topen}}{1+|\hat{\gscal{h}}_{\topen}|} -\frac{|\hat{\gscal{h}}_{\tclose}|\alpha_{\tclose}}{1+|\hat{\gscal{h}}_{\tclose}|}\right)\cdot {\rm op}_{\topen} \cdot 
 \zeta_{\topen}.
 \end{align}
\end{subequations}

As in \cite{chan2013algorithmic,orfanidis2020mathematical}, we consider pairs trading in which the  hedge ratio $\gscal{h}$ is allowed to changes over time. Accordingly, one is likely to have that $\hat{\gscal{h}}_{\topen} \neq \hat{\gscal{h}}_{\tclose}$, implying that the trader may open and close with different quantities of each of the assets. In Subsection \ref{ssec:discussion} we further discuss this property.

\subsection{Training Procedure}
\label{ssec:training}
The trading policy is stated in Algorithm~\ref{alg:Trading} for a given parameterization $\myVec{\theta}$. The tuning of these parameters is carried based on a data set comprised of a sequence of measured asset pairs $\mathcal{D}$ given in \eqref{eqn:Dataset}. To train $\myVec{\theta}$, we use a two-step training procedure, which first trains \acl{kn} separately from the trading task for stability, and then adapts the overall architecture based on the \ac{pnl}, thus tackling \ref{itm:pnl}. These two steps are detailed next. 

\subsubsection{Task-Ignorant Training}
This first step acts as a warm start. It tunes $\myVec{\theta}$ to be good initial values for training the the overall policy to maximize the \ac{pnl} in Step 2. To that aim, we train only \acl{knet} to the objective of optimizing the tracking of the state vector $\gvec{x_t}$ with stochastic gradient descent. 

In its original formulation in~\cite{revach2022kalmannet}, \acl{kn} is trained in a supervised manner,  i.e., using multiple trajectories of observations and their corresponding \acl{gt} states. Such a scheme cannot be applied here as the state $\gvec{x}_t$  comprises modelling parameters for which one cannot provide \acl{gt}. To overcome this challenge, we follow the unsupervised learning procedure proposed in \cite{revach2021unsupervised}, where the training loss is computed based on the predicted measurements, and not on the estimated state. The data set $\mathcal{D}$ is used to construct the sequence $\set{\gvec{F}_t, \gvec{g}_t, \gscal{y}_t}$, where
{the} prediction $\hat{\gscal{y}}_{t\given{t-1}}$ is computed using \eqref{eq:KFPredict}. The  resulting loss used for training $\myVec{\theta}$  is given by
%
\begin{equation}
\label{eqn:unsupervisedloss}
\mathcal{L}_{\mySet{D},1}(\myVec{\theta}) = \frac{1}{|\mySet{D}|}\sum_{(\alpha_t, \beta_t) \in \mySet{D}}\| \gscal{y}_t - \hat{\gscal{y}}_{t|t-1}\|^2.
\end{equation}
The  training (step 1) procedure is summarized as Algorithm~\ref{alg:TrainingStep1}.


\begin{algorithm}
\caption{Training Step 1 }
\label{alg:TrainingStep1} 
\SetKwInOut{Initialization}{Init}
\Initialization{Randomly initialize  $\myVec{\theta}$, step size $\eta_1>0$}
\SetKwInOut{Input}{Input} 
\Input{Initial estimate $\hat{\gvec{x}}_{0}=[\hat{\gscal{h}}_{0},\hat{\gscal{\mu}}_{0},\hat{\gscal{s}}_{0}]^T$;\\
Data set $\mathcal{D}$.}  
\For {each training epoch}
{Divide $\mySet{D}$ into $Q$ batches $\{\mySet{D}_q\}_{q=1}^Q$;\\
\For {each $\gscal{q}$ }{
\For {each $\gscal{t}$}{
Predict $\hat{\gscal{y}}_{t|t-1}$ via \eqref{eq:KNPredict}\;
Update estimate via   \eqref{eq:KNUpdated}\;
}
Compute $\mathcal{L}_{\mySet{D}_q,1}(\myVec{\theta})$ via \eqref{eqn:unsupervisedloss}\;
Update $\myVec{\theta} \leftarrow \myVec{\theta} - \eta_1 \nabla \mathcal{L}_{\mySet{D}_q,1}(\myVec{\theta})$ \;
}
}
\end{algorithm}

\subsubsection{Training End-to-End}
In this second and last step, we train \acl{kn} in an end-to-end manner based on the overall system task of pairs trading with full gradient descent. We adapt $\myVec{\theta}$ to maximize the  \ac{pnl}, namely, our loss function is
%
%
\begin{equation}
\label{eqn:E2Eloss}
\mathcal{L}_2(\myVec{\theta}) = -\sum_{i=0}^{N}{ r^{\varphi,\pi}_i}.
\end{equation}
In \eqref{eqn:E2Eloss},  $r^{\varphi,\pi}_i$ is the $i$th reward  given in \eqref{eq:total_reward_per_transaction} and ${N}$ is the number of active position taken.

A core challenge to optimize the trainable parameters with respect to  \eqref{eqn:E2Eloss}, stems from the non-differentiable nature of the policy in \eqref{eqn:OpenPolicy} and \eqref{eqn:ClosePolicyCum}. This prevents using conventional deep learning optimizers based on gradient descent. We overcome this challenge by approximating \eqref{eqn:OpenPolicy} and \eqref{eqn:ClosePolicyCum} with a {\em differentiable mapping during training}. Accordingly, we replace the unit step function with a parametric surrogates \cite{shlezinger2022deep}, $\unitstepp{\cdot}$, i.e., we compute the positions using \eqref{eqn:OpenPolicy} and \eqref{eqn:ClosePolicyCum}, while taking the gradient of \eqref{eqn:E2Eloss} assuming their surrogate approximations
\begin{equation}
\label{eqn:OpenPolicyS}
    \gscal{op}_t = \left(\unitstepp{-1 - \gscal{z}_t} -\unitstepp{\gscal{z}_t -1}\right) \cdot \unitstepp{\tau_{{\rm cp},t} -  \tau_{{\rm op},t}},
\end{equation}
and
 \begin{equation}
\label{eqn:ClosePolicyCumS}
   \gscal{cp}_t =    \unitstepp{-\gscal{z}_{t-1} \cdot \gscal{z}_{t}} \cdot (1 - \unitstepp{\tau_{{\rm cp},t} -  \tau_{{\rm op},t}}),
\end{equation}
respectively.
Following~\cite{shlezinger2022deep}, we set the surrogate approximation $\unitstepp{\cdot}$ to be the cumulative distribution function 
of a zero-mean Gaussian random variable with variance $\gamma^2$, i.e.,.
\begin{equation}
\label{eqn:surrogate}
    \unitstepp{x} = \int_{\tau=-\infty}^{x} \frac{1}{\sqrt{2\pi \gamma^2}}\exp\left\{-\frac{\tau^2}{2\gamma^2} \right\}d\tau.
\end{equation}
The differentiable nature and the simple derivative of the Gaussian function in \eqref{eqn:surrogate} enables computing the gradients of  \eqref{eqn:E2Eloss} using {backpropagation}, converting Algorithm~\ref{alg:Trading} into a discriminative trainable model~\cite{shlezinger2022discriminative}. The resulting second training step is summarized as Algorithm~\ref{alg:TrainingStep2}.


\begin{algorithm}
\caption{Training Step 2 }
\label{alg:TrainingStep2} 
\SetKwInOut{Initialization}{Init}
\Initialization{ Step size $\eta_2>0$}
\SetKwInOut{Input}{Input} 
\Input{Initial estimate $\hat{\gvec{x}}_{0}=[\hat{\gscal{h}}_{0},\hat{\gscal{\mu}}_{0},\hat{\gscal{s}}_{0}]^T$;\\ 
Data set $\mathcal{D}$; \\
Pre-trained parameters $\myVec{\theta}$ from Step 1.}  
\For {each training Epoch}{
\For {each $\gscal{t}$}{
Compute $\gscal{p}_t$ via  Algorithm~\ref{alg:Trading} with  $\myVec{\theta}$\;
}
Compute $\mathcal{L}_2(\myVec{\theta})$ via \eqref{eqn:E2Eloss}\;
Compute $\nabla \mathcal{L}_2(\myVec{\theta})$ with \eqref{eqn:OpenPolicy} and \eqref{eqn:ClosePolicyCum} approximated as \eqref{eqn:OpenPolicyS} and \eqref{eqn:ClosePolicyCumS}, respectively\;
Update $\myVec{\theta} \leftarrow \myVec{\theta} - \eta_2 \nabla \mathcal{L}_2(\myVec{\theta})$ \;}
\end{algorithm}

\subsection{Discussion}
\label{ssec:discussion}
\ac{kbpt} jointly exploits both the approximated \ac{ss} model representation in \eqref{eqn:OurSSModel} along with a principled augmentation of deep learning techniques.
This hybrid model-based/data-driven framework utilizes the strength of each one the approaches -- both classical techniques based on \ac{ss} models as well as data-driven deep learning tools. On one hand, we have  partial domain knowledge which we approximate using the \ac{pci} model of the system dynamics. This lets us maintain interpretability and transparency of the algorithm, which is crucial especially when one wants to invest based on such an algorithm. On the other hand, we use data and deep learning capabilities to elegantly overcome challenges \ref{itm:model}-\ref{itm:pnl}.

As empirically shown in Section~\ref{sec:emp_eval}, our hybrid policy allows to surpass the performance of other fully model-based or fully data-driven frameworks. The fact that \ac{kbpt} exploits the trainability of \acl{kn} and learns its tracking mapping from data based on the \ac{pnl} yields a tracking rule that does not necessarily give  the most accurate tracking, but is  useful for trading, which is our main objective. In addition, due to the interpretability of the framework, at each point in time, one can monitor the tracked latent state vector of the policy, and understand why each action was taken.

The integration of model-based deep learning techniques with financial decision making paves the way to a multitude of avenues for future exploration. Our derivation of Algorithm~\ref{alg:Trading} does not account for the transaction cost of each buy/sell action. As this is often a major factor in deciding if to perform such an action, one can potentially add to the \ac{pnl} function a penalty term for each transaction made. In that way, the second step of the training (Algorithm~\ref{alg:TrainingStep2}) will optimize with regards to the transaction costs. Nonetheless, it is empirically shown in Section~\ref{sec:emp_eval} that even without explicitly accounting for transaction cost, \ac{kbpt} still typically achieves its improved \ac{pnl} with lesser number of trades compared with model-based and data-driven benchmarks.
Another potential extension involves the usage of more than two assets. While we tackle the problem of pairs trading based on the statistical relation between two assets, one can generalize this approach and look for a relationship with multiple assets~\cite[Ch. 10]{feng2016signal}.

Another aspect of trading that can benefit from our hybrid model-based/data-driven design is  latency performance. Our experimental study reported in Section~\ref{sec:emp_eval} considers daily trading, which means an action is performed once a day. This framework can be implemented in high frequency trading, where a trader can perform hundreds of actions per day, and the latency of each step becomes a crucial factor. A core gain of such forms of model-based deep learning is their inference speed, which is often translated into much smaller latency compared with other deep learning models, and sometimes also compared with model-based processing, see, e.g., \cite{buchnik2023latent}. Likewise,  training  will be faster compared to fully data-driven approaches which is highly beneficial in a retraining regime. Therefore, we  expect our framework to be well suited for high-frequency trading, leaving this study for future work.

Lastly, as discussed following  \eqref{eq:assets_return}, when the hedge ratio changes over time, one needs to sell/buy more/less assets than  bought/sold when the position was opened. Such form of pairs trading of time-varying hedge ratio is not unique to our approach, and was also considered in, e.g., \cite{chan2013algorithmic,orfanidis2020mathematical}. A natural approach tackle this is have the trader posses a number of $\alpha$ and $\beta$ assets from the start, i.e., not just during open positions. 
Since the assets are co-integrated, the temporal variations in the hedge ratio are expected to be small, therefore the number of assets needed to begin with will be small. Moreover, one can encourage \acl{kn} to produce low variations in its tracked hedge by regularization, or even enforce it to be constant if starting with assets is to be avoided. 
%
%
We leave the exploration of this to future work.



%

\section{Empirical Evaluation}
\label{sec:emp_eval} 
%
In this section, we numerically evaluate \ac{kbpt}, comparing it with both \ac{ss} model-based and data-driven benchmarks. Our experimental study is comprised of four data sets (detailed in Subsection~\ref{ssec:ExpData}), and considers four benchmark trading policies (discussed in Subsection~\ref{ssec:benchmarks}). In our numerical evaluations we first assess the trading procedure and the individual contribution of each of its stages in Subsection~\ref{ssec:TrainingEval}, after which  we compare the   \ac{pnl} of \ac{kbpt} with the considered benchmarks in Subsection~\ref{ssec:Results}.

\subsection{Experimental Data}
\label{ssec:ExpData}
We consider four different data sets representing different asset pairs. 
Each data set is composed of in-sample data (training set)  and out-of-sample data (test set), where a sample consists of the daily opening prices of the two assets. The data sets are taken from \cite{yahoofianance} in such a way that the out-of-sample set starts one trading day after the end of the in-same set. 

We particularly use the following assets pairs
\begin{enumerate}
\item {\bf Swiss Frank - Euro (CHF - EURO)} - Consists of  2000 in-sample pairs, ending on 23/06/2019, and 944 out-of-sample pairs, starting from 24/06/2019.
\item {\bf Australian Dollar - South African Rand
 (AUD - ZAR)} - Consists of 2000 in-sample pairs, ending on 20/07/2017, and  1500 out-of-sample  pairs, starting from 21/07/2017.
 \item {\bf Canada ETF - Australia ETF Period A
 (EWC - EWA - A)} - Consists of  2000 in-sample pairs, ending on 29/01/2010, and  1500 out-of-sample  pairs, starting from 01/02/2010.
 \item {\bf Canada ETF - Australia ETF Period B
 (EWC - EWA - B)} - Consists of 2000 in-sample pairs, ending on 26/01/2017 and a 1500 out-of-sample  pairs, starting from 27/01/2010.
\end{enumerate}
These pairs were specifically selected due to the fact that they were found to be suitable for pairs trading, in the sense that they exhibit the \ac{ci} property, see e.g.,~\cite{chan2013algorithmic,feng2016signal,yu2022cointegration,orfanidis2020mathematical}.

\subsection{Benchmarks}
\label{ssec:benchmarks}
We compare \ac{kbpt} (Algorithm~\ref{alg:Trading}) with both model-based and data-driven  policies, using the following benchmarks:
\begin{enumerate}[label={\em B\arabic*}] 
\item \label{itm:KFCI} \ac{kf} with the \ac{ci} \ac{ss} model as detailed in Subsection \ref{ssec:CI_review}, followed by \ac{bb} policy in~\eqref{eqn:OpenPolicy} and \eqref{eqn:ClosePolicyCum}.  
\item \label{itm:KFPCI} \ac{kf} with the \ac{pci} \ac{ss} model as detailed in Subsection \ref{ssec:PCI_MODELs}, followed by  \ac{bb} policy in~\eqref{eqn:OpenPolicy} and \eqref{eqn:ClosePolicyCum}.  
\item \label{itm:KNetCI} Neural augmented \ac{kf}-\ac{bb} policy based on the \ac{ci} \ac{ss} model, i.e., Algorithm~\ref{alg:Trading} assuming the \ac{ss} model of Subsection \ref{ssec:CI_review} instead of our proposed \ac{ss} model of  detailed in Subsection \ref{ssec:PCI_MODELs}.
\item \label{itm:DDQN} A deep \ac{rl} trading policy based on the \ac{ddqn} architecture proposed in \cite{brim2020deep}, comprised of an input layer of 10 features, 2 fully connected layers of 50 neurons, and a 3 neurons output layer.
\end{enumerate}
Benchmarks \ref{itm:KFCI} and \ref{itm:KFPCI} represent  conventional model-based policy, wherein \ref{itm:KFPCI} is based on our proposed \ac{ss} formulation. Benchmark \ref{itm:DDQN} is purely data-driven, while \ref{itm:KNetCI} is a  model-based/data-driven approach, which is based on our  hybrid algorithm without incorporating the proposed \ac{ss} formulation.

\begin{figure}
  \centering
  \includegraphics[width=0.5\textwidth]{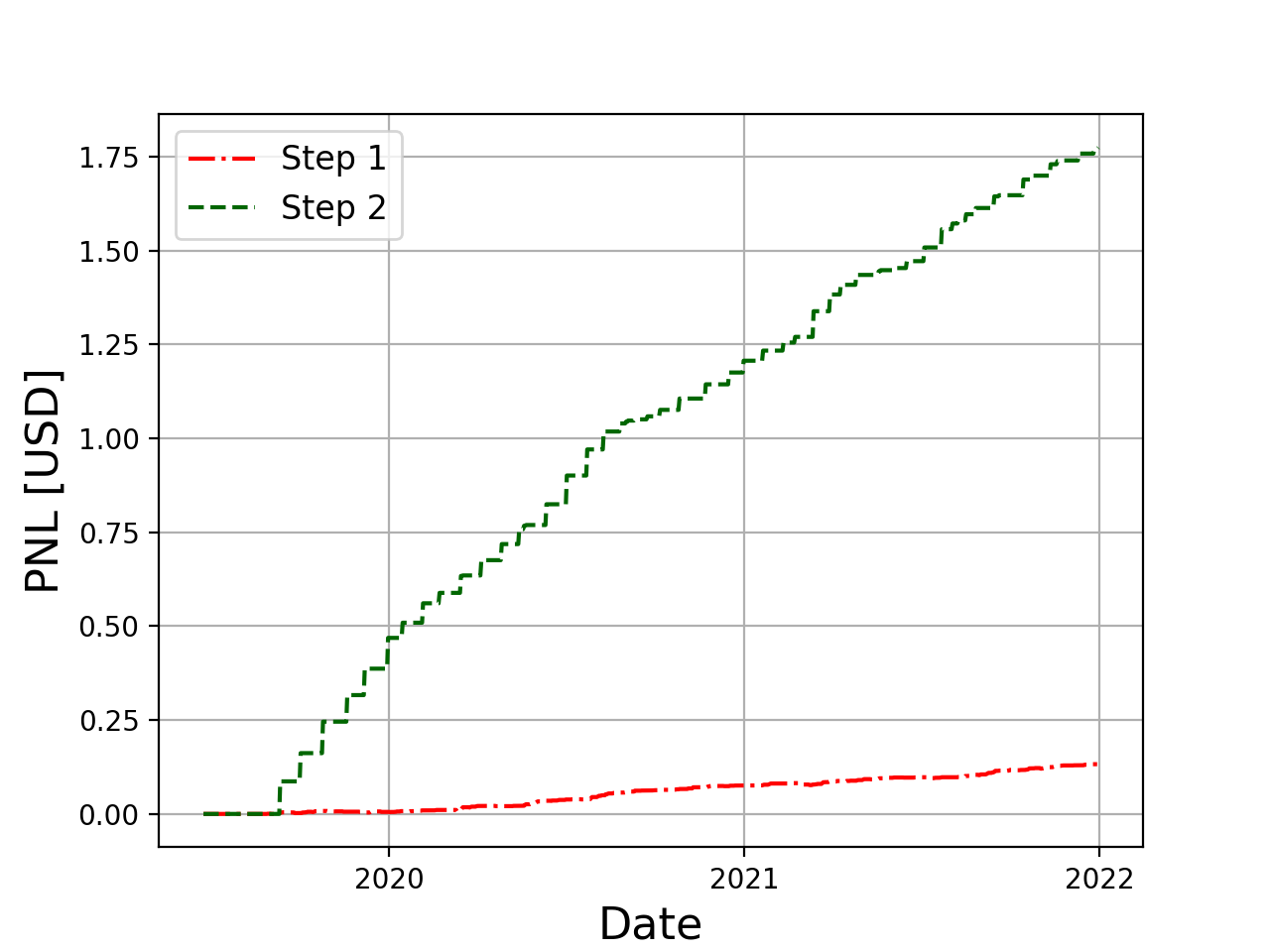}
  \caption{Comparison of the \ac{pnl} between the after the first training step to after the second on CHF - EURO pair}
  \label{fig:pnl_compare_steps}
\end{figure}

All data-driven algorithms, i.e., \ac{kbpt} as well as Benchmarks  \ref{itm:KNetCI} and \ref{itm:DDQN}, were trained using the Adam optimizer~\cite{kingma2014adam}, with  hyperparameters chosen by empirical trials\footnote{The source code and the complete set of hyperparameters used in our study is available at \url{https://github.com/KalmanNet/KBPT_TSP}.}. For KalmanNet, we used Architecture 2 of \cite{revach2022kalmannet} for the \ac{kg} \ac{rnn}. 
In the model-based policies, i.e. \ref{itm:KFCI} and \ref{itm:KFPCI}, we estimated the \ac{kf} parameters, e.g., the variances of the noise signals, from data. We intentionally used  the test set and not on the training set, giving a further advantage to the model-based approaches, in order to improve their performance and show that even under these conditions, they are still outperformed by our  our neural augmented design. Likewise was done on the \ac{ddqn} model of  \ref{itm:DDQN}, which we trained and evaluated  on the test set, giving it a clear advantage over our neural augmented design. Following~\cite{brim2020deep}, we trained it based on the instantaneous reward, allowing to obtain instantaneous loss for training.

\subsection{Evaluation of Training Procedure}
\label{ssec:TrainingEval}
We commence our experimental study by numerically evaluating  the contribution and usefulness of the different stages of the training procedure detailed in Subsection~\ref{ssec:training}. We particularly focus on their effect on the tracking performance and overall trading \ac{pnl}, as Step 1 (Algorithm~\ref{alg:TrainingStep1}) encourages accurate tracking of the assets, while its subsequent Step 2 (Algorithm~\ref{alg:TrainingStep2}) incorporates the \ac{pnl} objective.

\begin{figure}
  \centering
  \includegraphics[width=0.5\textwidth]{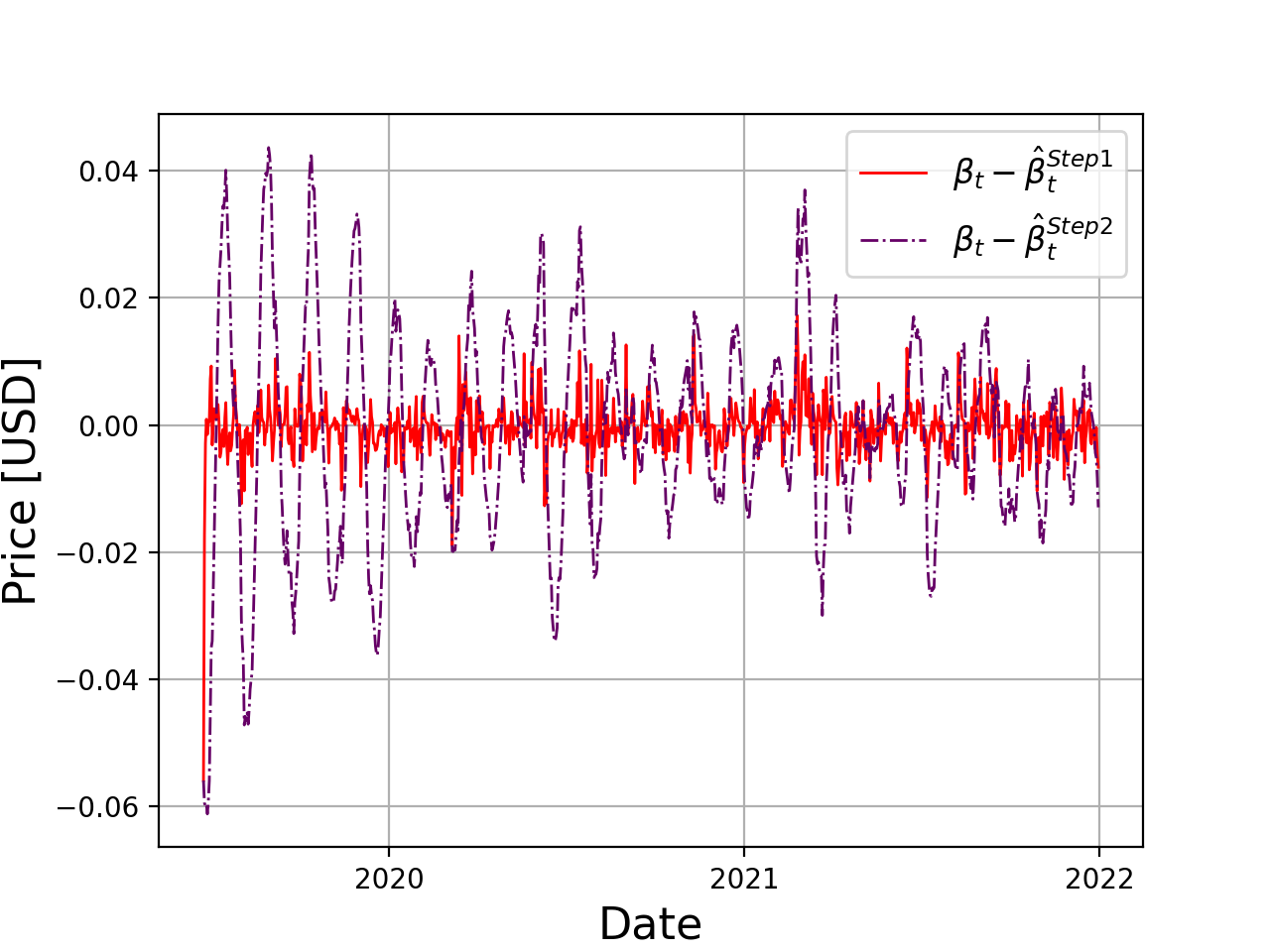}
  \caption{Comparison of the difference between the estimated and real observation (i.e. the second asset). $\hat{\beta}_t^{\rm Step1}$ and $\hat{\beta}_t^{\rm Step2}$ are the observations after the first and second step, respectively, and $\beta_t$ is the true asset price .}
  \label{fig:tracking_compare_steps}
\end{figure}

\begin{table}
\centering
\begin{tabular}{|c|c|c|c|}
\hline
{\textbf{Pair} }& & \textbf{Step 1} & \textbf{Step 2} \\
\hline
\multirow{2}{*}{CHF-EURO} & MSE [dB] & {\bf -47.60} & -35.64  \\
 & \ac{pnl} [USD] & 0.129 & {\bf 1.77 }\\
\hline
\multirow{2}{*}{AUD-ZAR} & MSE [dB] & {\bf-17.15}  & -3.28  \\
& \ac{pnl} [USD] & 0.008 & {\bf 9.19 }\\
\hline
\multirow{2}{*}{EWC-EWA-A} & MSE [dB]&  {\bf-12.8}  & -0.29   \\
& \ac{pnl} [USD] & -0.42 & {\bf 5.96}  \\
\hline
\multirow{2}{*}{EWC-EWA-B} & MSE [dB] & {\bf-11.61 } & -7.88  \\
& \ac{pnl} [USD] & 1.89 & {\bf 3.51}  \\
\hline
\end{tabular}
\vspace{0.2cm}
\caption{Comparison of average MSE in tracking the asset and final \ac{pnl} between the first and second  training steps}
\label{tabel:MSE_compare_steps}
\end{table}

In Fig.~\ref{fig:pnl_compare_steps} we evaluate \ac{kbpt} applied to the EURO-CHF pair, comparing its \ac{pnl} after trained only for tracking (Step 1) and after trained also for \ac{pnl} (Step 2). Observing Fig.~\ref{fig:pnl_compare_steps}, we clearly see that the addition of Step 2 notably boosts the \ac{pnl}, allowing to successfully cope with the challenge in incorporating this key consideration noted in \ref{itm:pnl}. It is emphasized though that training solely based on Step 2 (i.e., without preceding warm start via Step 1) makes learning much more challenging, and the training procedure often diverges.



To understand how training based on the \ac{pnl} rather than for state estimation affects the tracking performance, we depict in Fig.~\ref{fig:tracking_compare_steps} the difference between the estimated  $\hat{\beta}_t$ after each step and the real $\beta_t$ (which is the EURO price in USD). It is observed in Fig.~\ref{fig:tracking_compare_steps}  that (unsupervised) training based on the state estimation objective in Step 1 yields an accurate tracking of the assets. However, training based on the \ac{pnl} degrades the tracking performance, outputting a surrogate value which is most useful for trading in the sense of maximizing the \ac{pnl}. 
While Fig.~\ref{fig:tracking_compare_steps} focuses only on the EURO-CHF data, Table~\ref{tabel:MSE_compare_steps} reports that   tracking degradation (in \ac{mse}) is systematically converted into improved \ac{pnl} in all of pairs after the second training step. This study demonstrates the validity of our proposed two-step training procedure, and how it enables learning to track in a manner that contributes to the desired \ac{pnl} objective.

\begin{figure}
  \centering
  \includegraphics[width=0.5\textwidth]{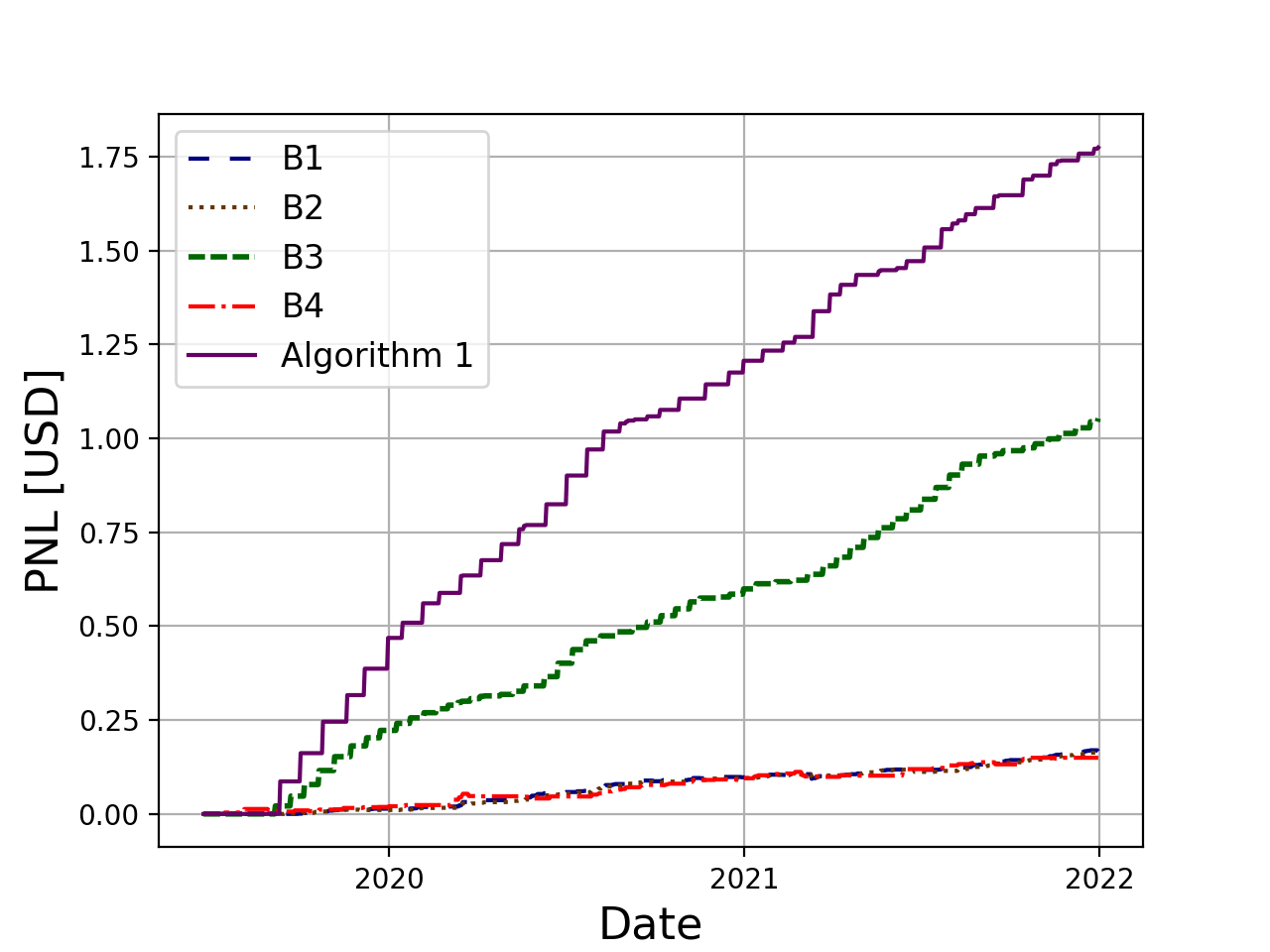}
  \caption{\ac{pnl} vs day index: Swiss Frank - Euro. }
  \label{fig:CHF_EURO_PNL}
\end{figure}

\subsection{Evaluation of \ac{pnl}}
\label{ssec:Results}
We proceed to evaluate the \ac{pnl} and its evolution in time as achieved by \ac{kbpt} and compared with benchmarks \ref{itm:KFCI}-\ref{itm:DDQN} 
 detailed in Subsection \ref{ssec:benchmarks}. We first consider the  CHF-EURO pair, and report the obtained \ac{pnl} versus day index $t$ in Fig.~\ref{fig:CHF_EURO_PNL}.  Observing Fig.~\ref{fig:CHF_EURO_PNL}, we note that both trading policies following our model-based deep learning approaches, i.e., Algorithm~\ref{alg:Trading} and \ref{itm:KNetCI}, outperform both the black-box data-driven \ac{ddqn} (\ref{itm:DDQN}) as well as  the fully model based benchmarks \ref{itm:KFCI} and \ref{itm:KFPCI}. Comparing the \ac{pnl} of Algorithm~\ref{alg:Trading} with \ref{itm:KNetCI}, as well as that of \ref{itm:KFPCI} with \ref{itm:KFCI}, it is consistently shown that our proposed \ac{pci} \ac{ss} model allows achieving improved \ac{pnl}, for both hybrid model-based/data-driven policies as well as purely model-based ones. 

 Moreover, we observe in Fig.~\ref{fig:CHF_EURO_PNL} (as well as in the following Figs.~\ref{fig:AUD_ZAR_PNL}-\ref{fig:EWC_EWA-B}) that the \ac{ddqn} approach, which relies on highly parameterized deep networks trained via \ac{rl}, achieves similar \ac{pnl} values to that of the much simpler \ac{kf}-\ac{bb} policy. It is noted that this is due to $(i)$ the \ac{ddqn} uses a constant hedge ratio; $(ii)$ \ac{rl} is notably facilitated when provided with instantaneous rewards, hence it is trained to provide policies based on an instantaneous reward. When the model-based \ac{kf}-\ac{bb} policy also determines its position based on the instantaneous rewards rather than the cumulative one, it is outperformed by the \ac{ddqn}, as reported in Table~\ref{tabel:Greedy_comparison}.

\begin{figure}
  \centering
  \includegraphics[width=0.5\textwidth]{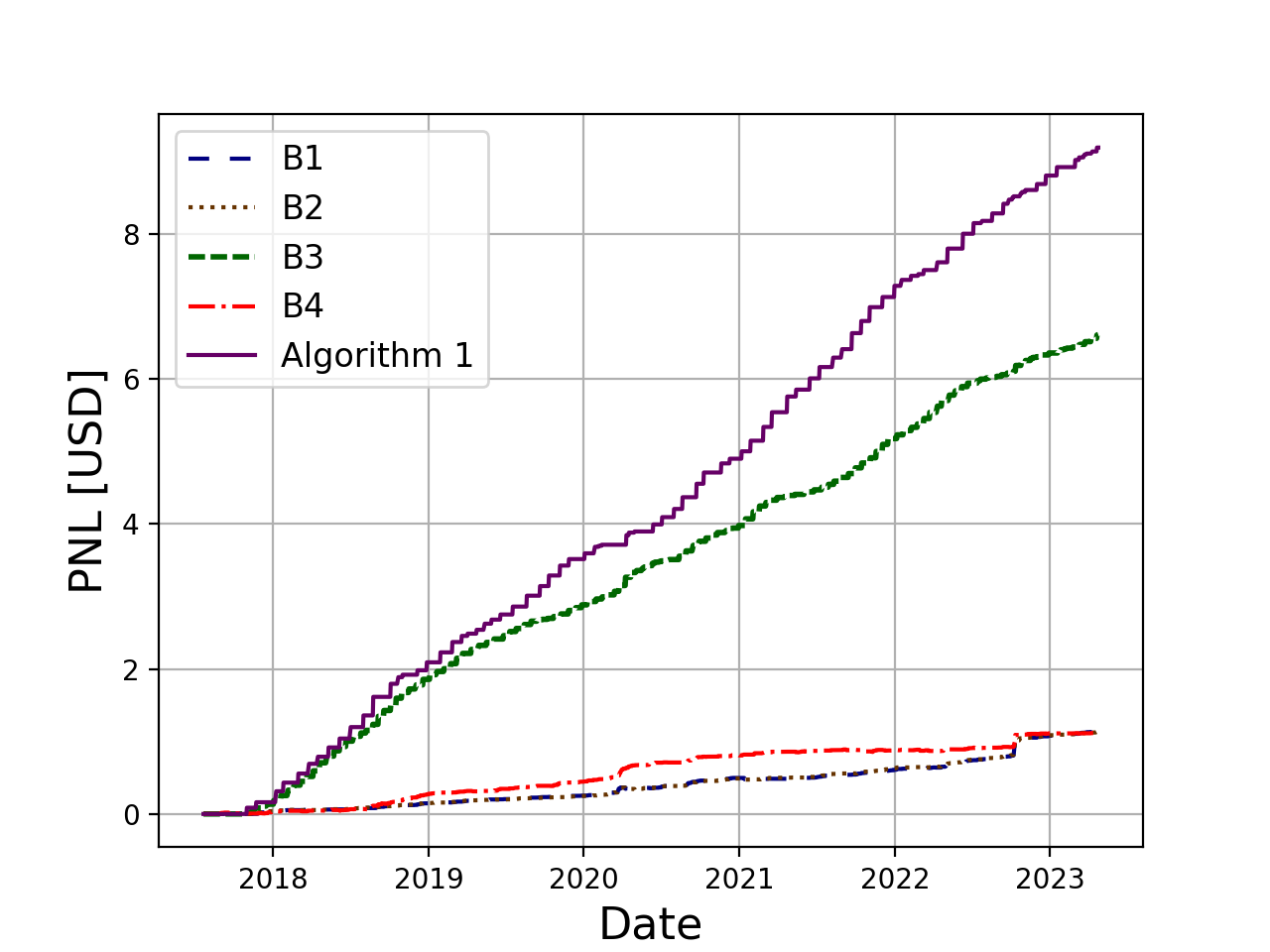}
  \caption{\ac{pnl} vs day index: Australian Dollar - South African Rand.}
  \label{fig:AUD_ZAR_PNL}
\end{figure}

\begin{table}
\centering
\begin{tabular}{|c|c|c|c|}
\hline
{\textbf{Pair} }& & \textbf{B1} & \textbf{B4} \\
\hline
{CHF-EURO} & \ac{pnl} [USD] & 0.051 & {\bf 0.149} \\
\hline
{AUD-ZAR} & \ac{pnl} [USD] & 0.27 &{\bf 1.11 }\\
\hline
{EWC-EWA-A} & \ac{pnl} [USD] & 0.62 & {\bf 1.09}  \\
\hline
{EWC-EWA-B}& \ac{pnl} [USD] & 0.43 & {\bf 1.57}  \\
\hline
\end{tabular}
\vspace{0.2cm}
\caption{Comparison of final \ac{pnl} with instantaneous reward}
\label{tabel:Greedy_comparison}
\end{table}
 

The findings reported in Fig.~\ref{fig:CHF_EURO_PNL} are not unique  to the CHF-EURO pair, and are also reproduced in Fig.~\ref{fig:AUD_ZAR_PNL}. There, we report the \ac{pnl} of a different pair which is also from the foreign exchange market, i.e., AUD - ZAR. We systematically observe both the superiority of a hybrid model-based/data-driven design, as well as the contribution of properly incorporating for expected variation profile of the spread via our \ac{ss} model formulated in \eqref{eqn:OurSSModel}.

\begin{figure}
  \centering
  \includegraphics[width=0.5\textwidth]{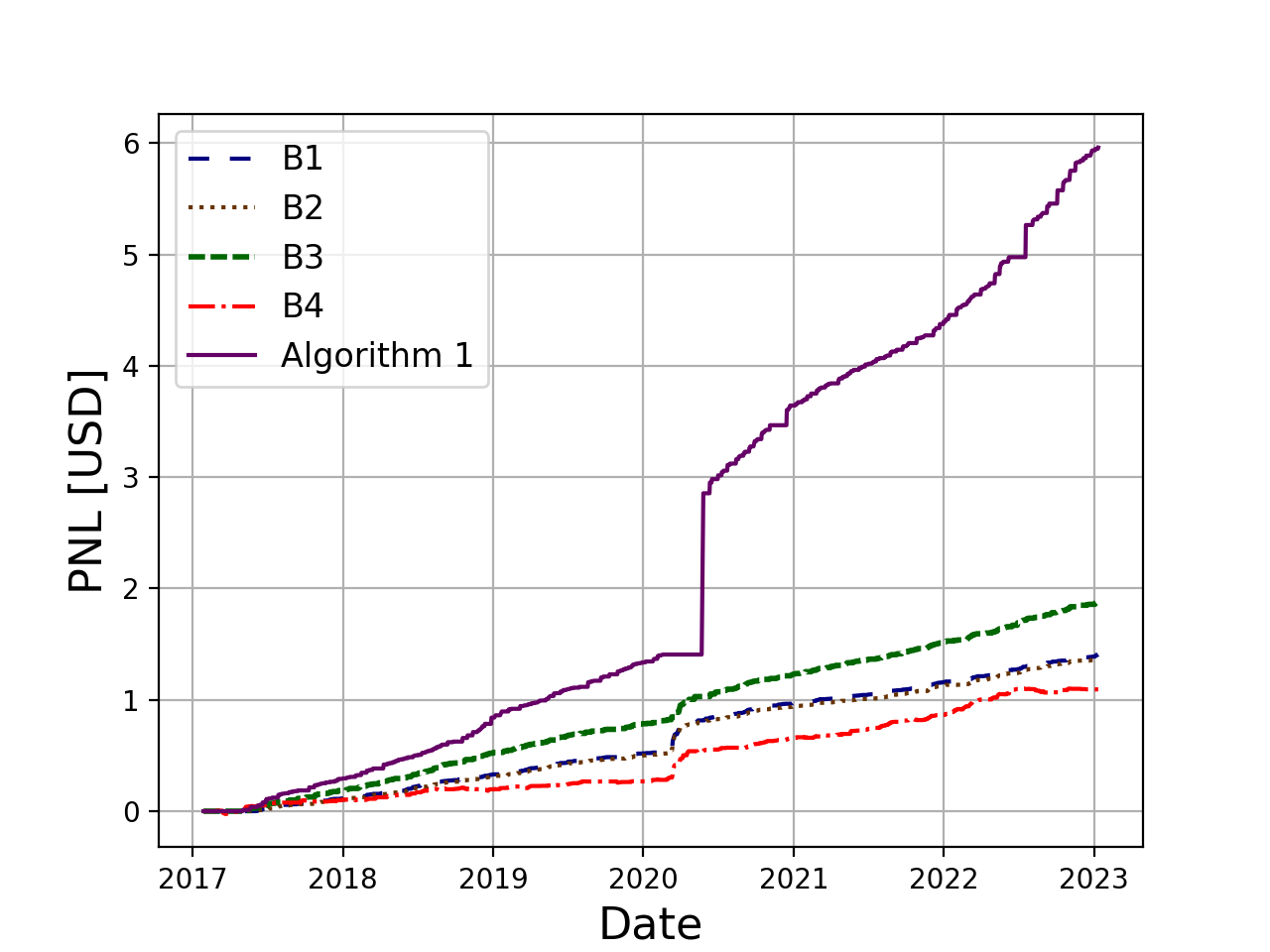}
  \caption{\ac{pnl} vs day index:  Canada ETF - Australia ETF Period A.}
  \label{fig:EWC_EWA-A}
\end{figure}

We demonstrate that the benefits of our design combining model-based deep learning with \ac{pci}-based \ac{ss} modelling are consistent over different markets as well. In Figs.~\ref{fig:EWC_EWA-A}-\ref{fig:EWC_EWA-B} we report the \ac{pnl} of EWC-EWA which are two ETLs, under two different time periods. These results showcase that the benefits of our design in terms of imported \ac{pnl} are  consistent over time and over markets that are known to be faithfully described as obeying some level of co-integration.

\begin{figure}
  \centering
  \includegraphics[width=0.5\textwidth]{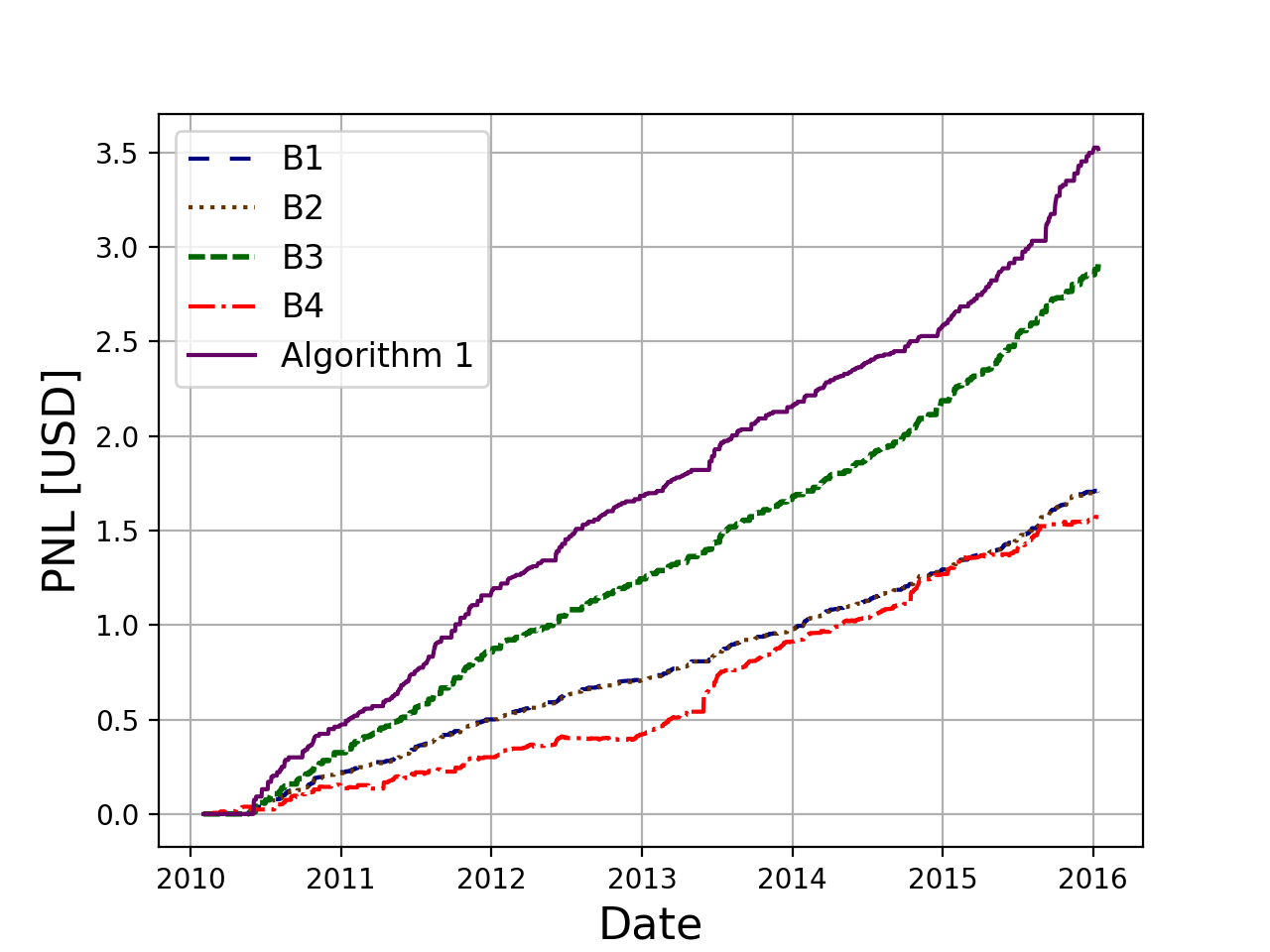}
  \caption{\ac{pnl} vs day index: Canada ETF - Australia ETF Period B. }
  \label{fig:EWC_EWA-B}
\end{figure}

The results reported so far evaluated trading in terms of \ac{pnl}. In practice, one is often interested in additional trading statistics, including the number of trades, annual return, mean return per trade, average holding time, and average time between returns~\cite{clegg2018pairs,han2023mastering}. 
In Table \ref{tabel:trade_stats} we summarize these trading statistics of the different pairs achieved by benchmarks \ref{itm:KFCI}-\ref{itm:DDQN} and our \ac{kbpt}. We show that our proposed trading policy performs less trades in the trading period but still achieves higher profit compared to that of other models. Although in our design we did not explicitly take into account the transaction cost by penalizing each trade transaction, we still obtained a trading policy that outperforms other trading policies and requires less transactions to do so.

\begin{table*}
\centering
\begin{tabular}{|c|c|c|c|c|c|c|}
\hline
{\textbf{Pair} }& {Metric }& \textbf{B1}& \textbf{B2} & \textbf{B3}& \textbf{B4}& \textbf{\ac{kbpt}}\\
\hline
\multirow{3}{*}{CHF-EURO} &Number of trades & 137 & 146 & 193 &557&{\bf 57}  \\
 & Annual return [\%] & 6.4 & 6.4 & 42 & 6 & {\bf70.}8  \\
 & Mean return per trade [\%] & 0.12 & 0.11 & 1.72 &0.02&{\bf3.11}  \\
 & Average holding time per trade [days] & 5.4 & 5.0 & 12.6 &1& 13.2 \\
 & Average time between returns [days] & 6.1 & 5.7 & 14.1 &1.6&15.0  \\
\hline
\multirow{3}{*}{AUD-ZAR} &Number of trades & 331 & 294 & 144 &1054&{\bf88} \\
& Annual return [\%] & 26.9 & 26.6 & 157 & 26.4 & {\bf218.8}  \\
 & Mean return per trade [\%] & 0.34 &  0.38 & 4.59 &0.10&{\bf10.44} \\
 & Average holding time per trade [days] & 3.6 & 4.1 & 8.9 &1 &14.9\\
 & Average time between returns [days]  & 4.2 & 4.8 & 9.8 &1.4 &16.4\\
\hline
\multirow{3}{*}{EWC-EWA-A} &Number of trades & 335 & 332 & 282 &1071&{\bf275}  \\
& Annual return [\%] & 33.3 & 32.6 & 44.5 & 25.9 & {\bf141.6}  \\
 & Mean return per trade [\%] & 0.39 & 0.41 & 0.66 &0.10&{\bf2.16}  \\
& Average holding time per trade [days] & 3.5 & 4.46 & 5.1 &1 &4.1\\
 & Average time between returns [days]  & 3.9 & 5.0 & 5.9 &1.3&5.2  \\
\hline
\multirow{3}{*}{EWC-EWA-B} &Number of trades & 340 & 342 & 415 &1126& {\bf278}  \\
& Annual return [\%] & 40.7 & 40.4 & 42.8 & 37.3 & {\bf83.5}  \\
 & Mean return per trade [\%] & 0.50 & 0.49 & 0.43&0.14&{\bf1.26} \\
 & Average holding time per trade [days] & 3.6 & 3.07 & 3.2 &1 &4.5\\
 & Average time between returns [days]  & 4.1 & 4.1 & 3.43 &1.3&5.1  \\
\hline
\end{tabular}
\vspace{0.2cm}
\caption{Trade statistics comparison between different models on different pairs. For the average time per trade  and the average time between returns metrics one can argue which value is preferred, hence no value is marked in bold fonts. }
\label{tabel:trade_stats}
\end{table*}

\section{Conclusions}
\label{ssec:conclusions}
We proposed \ac{kbpt}, a hybrid \acl{mb}/\acl{dd} trading policy that converts \ac{kf}-\ac{bb} pairs trading into a trainable architecture. It is based on an extended \ac{ss} model obtained from assuming partial co-integration. \ac{kbpt} utilizes the recent \acl{kn}  to learn to track while coping with the inherent mismatches in the underlying \ac{ss} model. Training is done in a two-step manner  combining unsupervised learning of \acl{kn} with  approximating the \ac{bb} policy with a differentiable mapping. Our empirical results show that the proposed policy notably improves \ac{kf}-\ac{bb} and deep \ac{rl} policies while preserving  interpretability, simplicity, and low latency. 
\bibliographystyle{IEEEtran}
\bibliography{IEEEabrv,refs}

\end{document}